\input ./style/arxiv-vmsta.cfg
\documentclass[numbers,compress]{vmsta}

\usepackage{eurosym}
\usepackage{longtable}

\volume{2}
\pubyear{2015}
\firstpage{67}
\lastpage{93}
\doi{10.15559/15-VMSTA25}

\startlocaldefs

\allowdisplaybreaks
\endlocaldefs

\begin{document}
\begin{frontmatter}

\title{Autoregressive approaches to import--export time series II: a
concrete case study}
\author{\inits{L.}\fnm{Luca}\snm{Di Persio}\corref{cor1}}\email
{dipersioluca@gmail.com}
\cortext[cor1]{Corresponding author.}
\author{\inits{C.}\fnm{Chiara}\snm{Segala}}


\address{Dept. Informatics, University of Verona, strada le Grazie 15,
37134, Italy}

\markboth{L. Di Persio, C. Segala}{Autoregressive approaches to
import--export time series II: a concrete case study}

\begin{abstract}
The present work constitutes the second part of a two-paper project
that, in particular, deals with an in-depth study of effective
techniques used in econometrics in order to make accurate forecasts in
the concrete framework of one of the major economies of the most
productive Italian area, namely the province of Verona. It is worth
mentioning that this region is indubitably recognized as the core of
the commercial engine of the whole Italian country. This is why our
analysis has a concrete impact; it is based on real data, and this is
also the reason why particular attention has been taken in treating the
relevant economical data and in choosing the right methods to manage
them to obtain good forecasts. In particular, we develop an approach
mainly based on vector autoregression where lagged values of two or
more variables are considered, Granger causality, and the stochastic
trend approach useful to work with the cointegration phenomenon.
\end{abstract}


\received{9 February 2015}
%
\revised{7 May 2015}
\accepted{12 May 2015}
\publishedonline{1 June 2015}
\end{frontmatter}

\begin{section}{Introduction}
In this second part of a two-paper project, we move from theory of
autoregressive, possibly multivalued, time series to the study of a
concrete framework. In particular, exploiting precious economic data
that the Commerce Chamber of Verona Province has put at our disposal,
we successfully applied some of the relevant approaches introduced in
\cite{rif9} to find dependencies between economic factors characterizing
the Province economy, then to make effective forecasts, very close to
the real behavior of studied markets. The present part of the project
is divided as follows:
first, we consider an AR-approach to Verona import--export time series,
then we provide a VAR model analysis of Verona relevant econometric
data taken from various web databases such as Coeweb, Stockview, and
Movimprese, and, within the last section, we compare such data with
those coming from the whole Italian scenario. We would like to
emphasize that all the theoretical background and related definitions
can be retrieved from \cite{rif9}.\vspace*{3pt}
\end{section}

\vspace*{-3pt}\section
{AR-approach to Verona import--export time series} \label{chapter3}\vspace*{-3pt} In what
follows, we shall apply techniques developed in previous sections to
analyze our main empirical problem of forecasting export and import
data for the Verona district, also using other variables such as active
enterprises. These applications are based on Istat data retrieved from
the database Coeweb.
\vspace*{3pt}\subsection{EXP} \label{EXP}
\begin{figure}[t]
\includegraphics{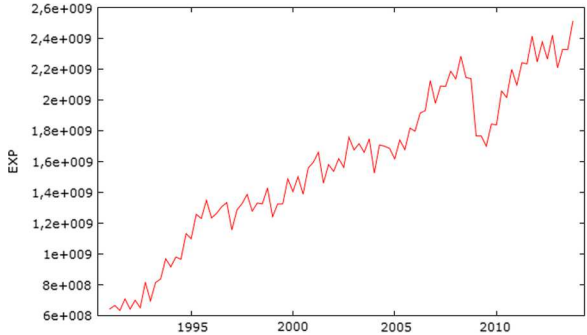}
\caption{Export of Verona}\label{exp}\vspace*{-6pt}
\end{figure}

We present a time series regression model in which the regressors are
past values of the dependent variable, namely the Export data. We use
92 observations of variable EXP, quarterly data from 1991 to 2013
expressed in Euros. Figure \ref{exp} shows the related time series.

Looking at Fig. \ref{exp}, we can see that the Verona export shows
relatively smooth growth, although this decreases during the years
2008--2011. Decline in exports is likely caused by economic crisis
broken out in Italy in those years. Although the curve may seem
apparently growing, it is also possible to notice that there are
periodic trends during the years under consideration. In fact, in the
fourth quarter of 1992, the curve has a significant growth, then
increases fairly linearly until about the second quarter of 1994, in
which one can recognize a new increasing period that slightly more
obvious than the previous one. This periodicity of 18 months can also
be seen in other parts of the curve, but not after the beginning of the
current economic crisis, where very likely there will be a structural
break. In\vadjust{\eject} order to test the goodness of our qualitative analysis based
on historical data, we used a software called GRETL, which is
particularly useful to perform statical analysis of time series. The
mean and standard deviation related to the quarter of this variable EXP
are respectively $ \mathit{Mean} = 1\,579\,900\,000 \  \mbox{\euro} \ \ \ \mathit{and} \ \ \
\mathit{Standard Deviation} = 499\,880\,000 \  \mbox{\euro}$, whereas the annual
mean for EXP is $ 1\,579\,900\,000 \times4 = 6\,319\,600\,000 \  \mbox{\euro}$.
The first seven autocorrelations of EXP are $ \rho_1 =
\mathit{corr}(\mathit{EXP}_t,\mathit{EXP}_{t-1})= 0.9718$, $\rho_2 = 0.9755$, $\rho_3 = 0.9450$,
$\rho_4 = 0.9523$, $\rho_5 = 0.9165$, $\rho_6 = 0.9242$, $\rho_7 =
0.8931$. Previous entries show that inflation is strongly positively
autocorrelated; in fact, the first autocorrelation is 0.97. The
autocorrelation remains large even at a lag of six quarters. This means
that an increase in export in one quarter tends to be associated with
an increase in the next quarter. Autocorrelation starts to decrease
from the lag of seventh quarters. In what follows, we report the output
obtained testing for autoregressive models according to an increasing
number of delays, from 1 to 6 delays, on the variable EXP, namely:

\bigskip
\noindent the AR(1) case: $\mathit{EXP} = 65\,090\,000 + 0.971606
\mathit{EXP}_{t-1}$
{\footnotesize
\begin{center}
%
\def\arraystretch{1.15}
\begin{tabular}{lr@{.}lr@{.}lr@{.}lr@{.}l}
&
\multicolumn{2}{c}{Coefficient} &
\multicolumn{2}{c}{Standard Error} &
\multicolumn{2}{c}{$t$-Statistic} &
\multicolumn{2}{c}{$p$-Value} \\[1ex]
const &
6&50900\textrm{e$+$007} &
2&35520\textrm{e$+$007} &
2&7637 &
0&0069 \\
$\mathit{EXP}_{t-1}$ &
0&971606 &
0&017392 &
55&8652 &
0&0000 \\[4.5pt]
\end{tabular}
\vspace{1ex}
\begin{tabular}{lrlr}
SER & \xch{1.17\textrm{e$+$08}}{1,17\textrm{e$+$08}} \\
$R^2$ & 0.944426 & Adjusted $R^2$ & 0.943802 \\
AIC & 3641.074 & BIC & 3646.096\\
\end{tabular}
\end{center}}
\noindent the AR(2) case: $\mathit{EXP} = 57\,965\,600 + 0.409313 \mathit{EXP}_{t-1} + 0.573763
\mathit{EXP}_{t-2}$
{\footnotesize
\begin{center}
%
\def\arraystretch{1.15}
\begin{tabular}{lr@{.}lr@{.}lr@{.}lr@{.}l}
&
\multicolumn{2}{c}{Coefficient} &
\multicolumn{2}{c}{Standard Error} &
\multicolumn{2}{c}{$t$-Statistic} &
\multicolumn{2}{c}{$p$-Value} \\[1ex]
const &
5&79656\textrm{e$+$007} &
2&92851\textrm{e$+$007} &
1&9794 &
0&0509 \\
$\mathit{EXP}_{t-1}$ &
0&409313 &
0&0920617 &
4&4461 &
0&0000 \\
$\mathit{EXP}_{t-2}$ &
0&573763 &
0&105188 &
5&4546 &
0&0000 \\
\end{tabular}

\vspace{1ex}
\begin{tabular}{lrlr}
SER & 97\,111\,006 \\
$R^2$ & 0.60913 & Adjusted $R^2$ & 0.960014 \\
AIC & 3568.804 & BIC & 3576.303\\[3pt]
\end{tabular}
\end{center}}
\noindent the AR(3) case: $
\mathit{EXP} = 54\,025\,100 + 0.618705 \mathit{EXP}_{t-1} + 0.726958 \mathit{EXP}_{t-2} -\break 0.366510
\mathit{EXP}_{t-3}$

{\footnotesize
\begin{center}

\def\arraystretch{1.15}
\begin{tabular}{lr@{.}lr@{.}lr@{.}lr@{.}l}
&
\multicolumn{2}{c}{Coefficient} &
\multicolumn{2}{c}{Standard Error} &
\multicolumn{2}{c}{$t$-Statistic} &
\multicolumn{2}{c}{$p$-Value} \\[1ex]
const &
5&40251\textrm{e$+$007} &
2&26874\textrm{e$+$007} &
2&3813 &
0&0195 \\
$\mathit{EXP}_{t-1}$ &
0&618705 &
0&109790 &
5&6353 &
0&0000 \\
$\mathit{EXP}_{t-2}$ &
0&726958 &
0&063352 &
11&4749 &
0&0000 \\
$\mathit{EXP}_{t-3}$ &
$-$0&366510 &
0&115843 &
$-$3&1639 &
0&0022 \\
\end{tabular}

\vspace{1ex}
\begin{tabular}{lrlr}
SER & 91\,264\,682 \\
$R^2$ & 0.964681 & Adjusted $R^2$ & 0.963435 \\
AIC & 3519.089 & BIC & 3529.044\\[4.5pt]
\end{tabular}
\end{center}}
\noindent the AR(4) case: $\mathit{EXP} =54\,498\,000 + 0.748057 \mathit{EXP}_{t-1} + 0.466614
\mathit{EXP}_{t-2} -\break 0.592869 \mathit{EXP}_{t-3}$

{\footnotesize
\begin{center}

\def\arraystretch{1.05}
\begin{tabular}{lr@{.}lr@{.}lr@{.}lr@{.}l}
&
\multicolumn{2}{c}{Coefficient} &
\multicolumn{2}{c}{Standard Error} &
\multicolumn{2}{c}{$t$-Statistic} &
\multicolumn{2}{c}{$p$-Value} \\[1ex]
const &
5&44980\textrm{e$+$007} &
2&43509\textrm{e$+$007} &
2&2380 &
0&0279 \\
$\mathit{EXP}_{t-1}$ &
0&748057 &
0&142495 &
5&2497 &
0&0000 \\
$\mathit{EXP}_{t-2}$ &
0&466614 &
0&075211 &
6&2041 &
0&0000 \\
$\mathit{EXP}_{t-3}$ &
$-$0&592869 &
0&156045 &
$-$3&7993 &
0&0003 \\
$\mathit{EXP}_{t-4}$ &
0&361048 &
0&065852 &
5&4827 &
0&0000 \\
\end{tabular}

\vspace{1ex}
\begin{tabular}{lrlr}
SER & 86\,223\,417 \\
$R^2$ & 0.967898 & Adjusted $R^2$ & 0.966351 \\
AIC & 3470.537 & BIC & 3482.924\\[4.5pt]
\end{tabular}
\end{center}}
\noindent the AR(5) case: $ \mathit{EXP} =56\,242\,200 + 0.870848 \mathit{EXP}_{t-1} + 0.247032
\mathit{EXP}_{t-2} -\break 0.417031 \mathit{EXP}_{t-3} \xch{+}{+ +} 0.648298 \mathit{EXP}_{t-4} - 0.372917
\mathit{EXP}_{t-5} $
{\footnotesize
\begin{center}

\def\arraystretch{1.05}
\begin{tabular}{lr@{.}lr@{.}lr@{.}lr@{.}l}
&
\multicolumn{2}{c}{Coefficient} &
\multicolumn{2}{c}{Standard Error} &
\multicolumn{2}{c}{$t$-Statistic} &
\multicolumn{2}{c}{$p$-Value} \\[1ex]
const &
5&62422\textrm{e$+$007} &
2&12088\textrm{e$+$007} &
2&6518 &
0&0096 \\
$\mathit{EXP}_{t-1}$ &
0&870848 &
0&135548 &
6&4246 &
0&0000 \\
$\mathit{EXP}_{t-2}$ &
0&247032 &
0&096569 &
2&5581 &
0&0124 \\
$\mathit{EXP}_{t-3}$ &
$-$0&417031 &
0&178982 &
$-$2&3300 &
0&0223 \\
$\mathit{EXP}_{t-4}$ &
0&648298 &
0&105669 &
6&1352 &
0&0000 \\
$\mathit{EXP}_{t-5}$ &
$-$0&372917 &
0&119834 &
$-$3&1119 &
0&0026 \\
\end{tabular}

\vspace{1ex}
\begin{tabular}{lrlr}
SER & 80\,872\,743 \\
$R^2$ & 0.970976 & Adjusted $R^2$ & 0.969185 \\
AIC & 3420.938 & BIC & 3435.733\\
\end{tabular}
\end{center}}
\noindent and the AR(6) case:
\begin{align}
\label{AR(6)} %
\mathit{EXP} &=55434600 + 1.01304 \mathit{EXP}_{t-1} + 0.00610464 \mathit{EXP}_{t-2} - 0.251406 \mathit{EXP}_{t-3}\nonumber\\
&\quad + 0.542831 \mathit{EXP}_{t-4} - 0.737681 \mathit{EXP}_{t-5} + 0.408104 \mathit{EXP}_{t-6}
\end{align}

{\footnotesize
\begin{center}

\def\arraystretch{1.05}
\begin{tabular}{lr@{.}lr@{.}lr@{.}lr@{.}l}
&
\multicolumn{2}{c}{Coefficient} &
\multicolumn{2}{c}{Standard Error} &
\multicolumn{2}{c}{$t$-Statistic} &
\multicolumn{2}{c}{$p$-Value} \\[1ex]
const &
5&54346\textrm{e$+$007} &
2&23371\textrm{e$+$007} &
2&4817 &
0&0152 \\
$\mathit{EXP}_{t-1}$ &
1&01304 &
0&12541 &
8&0777 &
0&0000 \\
$\mathit{EXP}_{t-2}$ &
0&006105 &
0&107043 &
0&0570 &
0&9547 \\
$\mathit{EXP}_{t-3}$ &
$-$0&251406 &
0&131646 &
$-$1&9097 &
0&0598 \\
$\mathit{EXP}_{t-4}$ &
0&542831 &
0&116130 &
4&6743 &
0&0000 \\
$\mathit{EXP}_{t-5}$ &
$-$0&737681 &
0&104151 &
$-$7&0828 &
0&0000 \\
$\mathit{EXP}_{t-6}$ &
0&408104 &
0&089469 &
4&5614 &
0&0000 \\
\end{tabular}

\vspace{1ex}
\begin{tabular}{lrlr}
SER & 75\,057\,009 \\
$R^2$ & 0.974384 & Adjusted $R^2$ & 0.972438 \\
AIC & 3369.763 & BIC & 3386.943\\
\end{tabular}
\end{center}
} We estimate the AR order of our autoregression related to obtained
numerical results using both BIC and AIC information \xch{criteria (see Table~\ref{criteria}).}{criteria:}

\begin{table}[h]
\caption{BIC, \xch{AIC}{AI.}, Adjusted $R^2$, and SER for the six AR
models}\label{criteria}
\begin{tabular}{ccccc}
\hline
$p$ & $\mathit{BIC}(p)$ & $\mathit{AIC}(p)$ &
$\mathit{Adjusted}\ R^2 (p)$ & $\mathit{SER}(p)$ \\
\hline
1 & 3646.096 & 3641.074 & 0.943802 & 117000000 \\
2 & 3576.303 & 3568.804 & 0.960014 & 97111006 \\
3 & 3529.044 & 3519,089 & 0.963435 & 91264682 \\
4 & 3482.924 & 3470.537 & 0.966351 & 86223417 \\
5 & 3435.733 & 3420.938 & 0.969185 & 80872743 \\
6 & 3386.943 & 3369.763 & 0.972438 & 75057009 \\
\hline
\end{tabular}
\end{table}

Both BIC and AIC are the smallest in the AR(6) model (from the seventh
delay onwards the criteria begin to increase); we conclude that the
best estimate of the lag length is 6, hence supporting our qualitative
analysis. Previous data from Table \ref{criteria} indicate that as the
number of lags increases, the $\mathit{Adjusted}\ R^2$ increases, and the SER
decreases. $R^2, \ \mathit{Adjusted}\,R^2, \mathit{\,and\,SER} $ measure how well the OLS
estimate of the multiple regression line describes the data. The
standard error of the regression (SER) estimates the standard deviation
of the error term, and thus, it is a measure of spread of the
distribution of a~variable~$Y$ around the regression line. The
regression $R^2$ is the fraction of the sample variance of $Y$
explained by (or predicted by) the regressors,~the $R^2$ increases
whenever a regressor is added, unless the estimated coefficient on the
added regressor is exactly zero. An increase in the $R^2$ does not mean
that adding a variable actually improves the fit of the model, so the
$R^2$ gives an inflated estimate of how well the regression fits the
data. One way to correct this is to deflate or reduce the $R^2$ by
some factor, and this is what the $\mathit{Adjusted}\,R^2$ does, which is a
modified version of $R^2$ that does not necessarily increase when a new
regressor is added. As seen by numerical output in
Table~\ref{criteria}, the increase in $\mathit{Adjusted}\,R^2$ is large from one
to two lags, smaller from two to three, and quite small from three to
four and in the~next lags. Exploiting the results obtained for the
AIC/BIC analysis, we can determine how large the increase in the
$\mathit{Adjusted}\,R^2$ must be to justify including the additional lag.~In the
AR(6) model of Eq.~\eqref{AR(6)}, the coefficients of $\mathit{EXP}_{\!t-1}$,
$\mathit{EXP}_{\!t-4}$, $\mathit{EXP}_{\!t-5}$, and $\mathit{EXP}_{\!t-6}$ are statically significant at
the $1\%$ significance level because their $p$-value is less than
$0{.}01$, and the $t$-statistic exceeds the critical value. The
constant, however, is statically significant at the $5\%$ significance.
The coefficient of $\mathit{EXP}_{t-3}$ is statically significant at the $10\%$
significance, and the coefficient of $\mathit{EXP}_{t-2}$ is not statically
significant. In particular, the $95\%$ confidence intervals for these
coefficient are as follows:\looseness=-1

{\footnotesize
\begin{center}
\begin{tabular}{rr@{.}lr@{.}lr@{.}l}
Variable%
& \multicolumn{2}{c}{Coefficient}%
& \multicolumn{4}{c}{$95\%$ Confidence Interval}\\[1ex]
const & 5&54346\textrm{e$+$007} & 1&09738\textrm{e$+$007} & 9&98955\textrm
{e$+$007}\\
$\mathit{EXP}_{t-1}$ & 1&01304 & 0&76341 & 1&26266\\
$\mathit{EXP}_{t-2}$ & 0&006105 & $-$0&206959 & 0&219168\\
$\mathit{EXP}_{t-3}$ & $-$0&251406 & $-$0&513441 & 0&010627\\
$\mathit{EXP}_{t-4}$ & 0&542831 & 0&311680 & 0&773981\\
$\mathit{EXP}_{t-5}$ & $-$0&737681 & $-$0&944989 & $-$0&530374\\
$\mathit{EXP}_{t-6}$ & 0&408104 & 0&230022 & 0&586187\\
\end{tabular}
\end{center}
}

\begin{table}[t]
\caption{Large-sample critical values of the augmented Dickey--Fuller statistic}\label{ADF}
\begin{tabular}{cccc}
\hline
$\mathit{Deterministic \ Regressors}$ & ${10 \ \%}$ &
${5 \ \%}$ & ${1 \ \%}$ \\
\hline
Intercept only & --2.57 & --2.86 & --3.43 \\
Intercept and time trend & --3.12 & --3.41 & --3.96 \\
\hline
\end{tabular}\vspace*{-6pt}
\end{table}

In order to check whether the EXP variable has a trend component or
not, we test the null hypothesis that such a trend actually exists
against the alternative EXP being stationary, by performing the ADF
test for a unit autoregressive root. Large-sample critical values of
the augmented Dickey--Fuller statistic yield
the following ADF regression with six lags of $\mathit{EXP}_t$, where the
subscript $t$ indicates a particular quarter considered:
\begin{align}
\label{ADFeq} %
 \widehat{\Delta \mathit{EXP}}_t &= 55\,434\,600 + \delta \mathit{EXP}_{t-1} + \gamma_1 \Delta
\mathit{EXP}_{t-1} + \gamma_2 \mathit{EXP}_{t-2}\nonumber\\
& \quad + \gamma_3 \Delta \mathit{EXP}_{t-3} + \gamma_4 \Delta
\mathit{EXP}_{t-4} + \gamma_5 \Delta \mathit{EXP}_{t-5} +
\gamma_6 \Delta \mathit{EXP}_{t-6}.\vadjust{\goodbreak}
\end{align}
The ADF $t$-statistic is the $t$-statistic testing the hypothesis that
the coefficient on $\mathit{EXP}_{t-1}$ is zero; this is $t = -1.23$. From Table
\ref{ADF}, the 5\% critical value is $-2.86$. Because the ADF statistic
of $-1.23$ is less negative than $-2.86$, the test does not reject the
null hypothesis at the 5\% significance level. Based on the regression
in Eq.~\eqref{ADFeq}, we therefore cannot reject the null hypothesis
that export has a unit autoregressive root, that is, that export
contains a stochastic trend, against the alternative that it is
stationary. If instead the alternative hypothesis is that $Y_t$ is
stationary around a deterministic linear trend, then the ADF
$t$-statistic results in $t = -4.07$, which is less than \xch{$-3{.}41$}{$-3,41$} (from
Table \ref{ADF}). Hence, we can reject the null hypothesis that export
has a unit autoregressive root. We proceed with a test QLR, which
provides a way to check whether the export curve has been stable in the
period from 1993 to 2010. Specifically, we focus on whether there have
been changes in the coefficients of the lagged values of export and of
the intercept in the AR(6) model specification in Eq.~\eqref{AR(6)}
containing six lags of $\mathit{EXP}_t$. The Chow F-statistics (see, e.g.,
\cite[Sect.~5.3.3]{rif1}) tests the hypothesis that the intercept and
the coefficients of $\mathit{EXP}_{t-1} , \ldots, \mathit{EXP}_{t-6}$ in
Eq.~\eqref{AR(6)} are constant against the alternative that they break
at a given date for breaks in the central 70\% of the sample. The
F-statistic is computed for break dates in the central 70\% of the
sample because for the large-sample approximation to the distribution
of the QLR statistic to be a good one, the subsample endpoints cannot
be too close to the beginning or to the end of the sample, so we decide
to use 15\% trimming, that is, to set $\tau_0 = 0.15T$ and $\tau_1 =
0.85T$ (rounded to the nearest integer). Each F-statistic tests seven
restrictions. Restrictions on the coefficients equaled to zero under
the null hypothesis (see \cite[Sect.~2.4]{rif9}), and since in our case
we have the coefficients of the six delays and the intercept, we get
seven restrictions. The largest of these F-statistics is \xch{13.96}{13,96}, which
occurs in 2010:I (the first quarter of 2010); this is the QLR
statistic. The critical value for seven restrictions \xch{is presented in Table~\ref{QLR}.}{is as
follows:}

\begin{table}[t]
\caption{Critical values of QLR statistic with 15\% truncation}\label{QLR}
\begin{tabular}{cccc}
\hline
$\mathit{Number \ of \ restrictions}$ & ${10 \ \%}$ &
${5 \ \%}$ & ${1 \ \%}$ \\
\hline
7 & \xch{2.84}{2,84} & \xch{3.15}{3,15} & \xch{3.82}{3,82} \\
\hline
\end{tabular}
\end{table}

The previously reported values indicate that the hypothesis of stable
coefficients is rejected at the 1\% significance level. Thus, there is
an evidence that at least one of these seven coefficients changed over
the sample. These results also confirm the assumptions that we made
earlier since the year 2010 coincides with an increasing import of the
financial crisis before arriving at a partial economic recovery. A
forecast of Verona export in 2014:I using data through 2013:IV can be
then based on our established AR(6) model of export, which
gives
\begin{align*}
\mathit{EXP}& =55\,434\,600 + 1.01304 \mathit{EXP}_{t-1} + 0.00610464 \mathit{EXP}_{t-2} -
0.251406 \mathit{EXP}_{t-3}
\\
&\quad{}+0.542831 \mathit{EXP}_{t-4} - 0.737681 \mathit{EXP}_{t-5} + 0.408104
\mathit{EXP}_{t-6}.
\end{align*}
Therefore, substituting the values of export into each of the four
quarters of 2013, plus the two last quarters of 2012, we have
\begin{align*}
\widehat{\mathit{EXP}}_{2014:I | 2013:IV} &=55\,434\,600 + \xch{1.013}{1,013}\mathit{EXP}_{2013:\mathrm{IV}} + \xch{0.006}{0,006} \mathit{EXP}_{2013:\mathrm{III}}\\
& \quad - \xch{0.251}{0,251} \mathit{EXP}_{2013:\mathrm{II}} + \xch{0.543}{0,543} \mathit{EXP}_{2013:\mathrm{I}}\\
& \quad - \xch{0.738}{0,738} \mathit{EXP}_{2012:\mathrm{IV}} + \xch{0.408}{0,408} \mathit{EXP}_{2012:\mathrm{III}}\\[1pt]
&= 55\,434\,600 + \xch{1.013}{1,013} \times2\,511\,098\,163 + \xch{0.006}{0,006} \times2\,326\,958\,115\\[1pt]
& \quad - \xch{0.251}{0,251} \times2\,329\,551\,351 + \xch{0.543}{0,543} \times2\,209\,212\,521\\[1pt]
& \quad - \xch{0.738}{0,738} \times2\,420\,606\,501 + \xch{0.408}{0,408} \times 2\,265\,903\,940 \\[1pt]
&\cong2\,366\,137\,617 \ \mbox{\euro} ,
\end{align*}

\noindent so that, for 2014:II, we obtain
\begin{align*}
\widehat{\mathit{EXP}}_{2014:\mathrm{II} | 2014:\mathrm{I}} &=55\,434\,600 + \xch{1.013}{1,013}\mathit{EXP}_{2014:\mathrm{I}} + \xch{0.006}{0,006} \mathit{EXP}_{2013:\mathrm{IV}}\\[1pt]
& \quad - \xch{0.251}{0,251} \mathit{EXP}_{2013:\mathrm{III}} + \xch{0.543}{0,543} \mathit{EXP}_{2013:\mathrm{II}}\\[1pt]
& \quad - \xch{0.738}{0,738} \mathit{EXP}_{2012:\mathrm{I}} + \xch{0.408}{0,408} \mathit{EXP}_{2012:\mathrm{IV}}\\[1pt]
&= 55\,434\,600 + \xch{1.013}{1,013} \times2\,366\,137\,617 + \xch{0.006}{0,006} \times2\,511\,098\,163\\[1pt]
& \quad - \xch{0.251}{0,251} \times2\,326\,958\,115 + \xch{0.543}{0,543} \times2\,329\,551\,351\\[1pt]
& \quad - \xch{0.738}{0,738} \times2\,209\,212\,521 + \xch{0.408}{0,408} \times 2\,420\,606\,501 \\[1pt]
&\cong 2\,505\,454\,123 \ \mbox{\euro} ,
\end{align*}
and forecasts for all 2014 quarters are as follows:
\begin{center}
{\footnotesize
\begin{tabular}{ccc}
\hline
$\mathit{Quarter}$ & $\mathit{Forecast}$ & $\mathit{Error}$
\\
\hline
2014:I & 2\,366\,130\,000 & 75\,057\,000 \\

2014:II & 2\,505\,450\,000 & 106\,841\,000 \\

2014:III & 2\,422\,950\,000 & 131\,981\,000 \\

2014:IV & 2\,527\,660\,000 & 145\,016\,000 \\
\hline
\end{tabular}}\vspace*{3pt}
\end{center}

\begin{figure}[b!]
\includegraphics[scale=1.05]{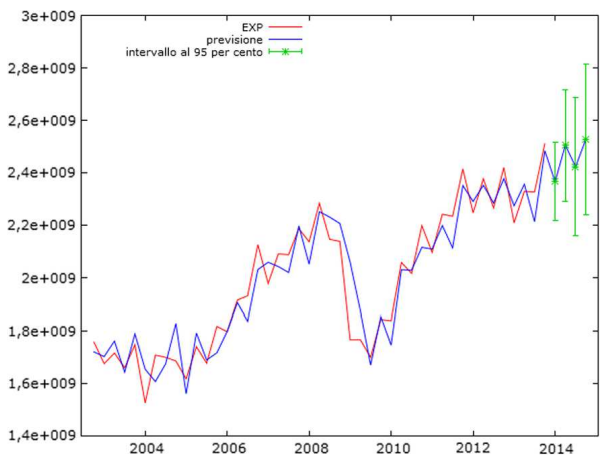}
\caption{Forecasts of Verona export}\label{exprev}
\end{figure}

It is worth mentioning that the forecast error increases as the
number of considered quarters increases. Figure \ref{exprev} shows,
through a graph, forecasts since 2002 in sample and forecasts for 2014,
highlighting the confidence intervals.

\subsection{$\Delta \mathit{EXP}$}

\begin{figure}[t]
\includegraphics{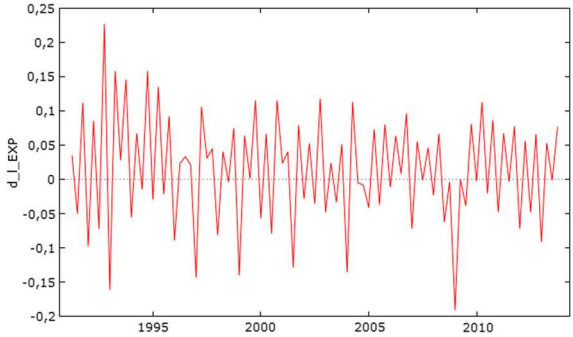}
\caption{Rate of growth in exports}\label{deltaexp}
\end{figure}

It is also useful to analyze the time series of the growth rate in
exports that we denoted by $\Delta \mathit{EXP}$. Economic time series are often
analyzed after computing their logarithms or the changes in their
logarithms. One reason for this is that many economic series exhibit
growth that is approximately exponential, that is, over the long run,
the series tends to grow by a certain percentage per year on average,
and hence the logarithm of the series grows approximately linearly.
Another reason is that the standard deviation of many economic time
series is approximately proportional to its level, that is, the
standard deviation is well expressed as a percentage of the level of
the series; hence, if this is the case, the standard deviation of the
logarithm of the series is approximately constant. It follows that it
turns to be convenient to work with the variable $\Delta \mathit{EXP}_t =
\ln(\mathit{EXP}_t) - \ln(\mathit{EXP}_{t-1}) $. Taking into account the data shared in
Fig.~\ref{deltaexp}, we retrieve the following information:
\begin{displaymath}
\mathit{Mean \ on \ a \ quarterly \ basis}= 0.014958 = 1.49 \%
\end{displaymath}
\begin{displaymath}
\mathit{Standard \ Deviation \ on \ a \ quarterly \ basis}= 0.079272 = 7.93 \%
\end{displaymath}
\begin{displaymath}
\mathit{Average \ Growth \ Rate \ on \ a \ yearly \ basis}
= 0.014958 \times4 = 0.059832 = 5.98 \%
\end{displaymath}
The first four autocorrelations of $\Delta \mathit{EXP}$ are $ \rho_1 =
-0.6133$, $\rho_2 = 0.5698$, $\rho_3 = -0.6100$, $\rho_4 = 0.7029 $.

Even if it might seem contradictory that the level of export is
strongly positively correlated but its change is negatively correlated,
we have to consider that such values measure different things. The
strong positive autocorrelation in export reflects the long-term trends
in export; in contrast, the negative autocorrelation of the change of
export means that, on average, an increase in export in one quarter is
associated with a decrease in export in the next one. Analogously to
what we have seen in \xch{Section}{Subsection}~\ref{EXP}, we perform an AIC/BIC
analysis for $\Delta \mathit{EXP}$ obtaining that the best choice for the lag
lay is 4, so that we have
\begin{align*}
\Delta \mathit{EXP} &= 0.0128189 - 0.173627 \Delta \mathit{EXP}_{t-1} + 0.0996175 \Delta
\mathit{EXP}_{t-2}
\\
&\quad{}- 0.189882 \Delta \mathit{EXP}_{t-3} + 0.416414 \Delta
\mathit{EXP}_{t-4}.
\end{align*}
\begin{center}

\vspace{1em} {\footnotesize
\begin{tabular}{lr@{.}lr@{.}lr@{.}lr@{.}l}
&
\multicolumn{2}{c}{Coefficient} &
\multicolumn{2}{c}{Standard Error} &
\multicolumn{2}{c}{$t$-Statistic} &
\multicolumn{2}{c}{$p$-Value} \\[1ex]
const &
0&0128189 &
0&0077887 &
1&6458 &
0&1036 \\
$\Delta \mathit{EXP}_{t-1}$ &
$-$0&173627 &
0&119987 &
$-$1&4470 &
0&1517 \\
$\Delta \mathit{EXP}_{t-2}$ &
0&099618 &
0&100542 &
0&9908 &
0&3247 \\
$\Delta \mathit{EXP}_{t-3}$ &
$-$0&189882 &
0&096363 &
$-$1&9705 &
0&0522 \\
$\Delta \mathit{EXP}_{t-4}$ &
0&416414 &
0&094464 &
4&4081 &
0&0000 \\[4.5pt]
\end{tabular}
\vspace{1ex}
\begin{tabular}{lrlr}
SER & 0.052736 \\
$R^2$ & 0.576787 & Adjusted $R^2$ & 0.556142 \\
AIC & --260.2402 & BIC & --247.9107\\
\end{tabular}}
\end{center}

\begin{figure}[t]
\includegraphics{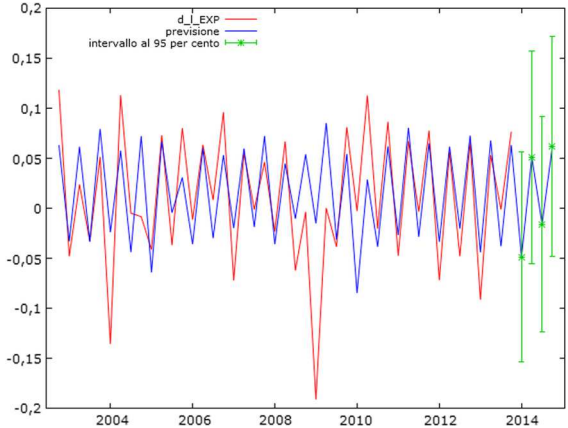}
\caption{Forecasts of ${\Delta \mathit{EXP}}$}\label{deltaexprev}
\end{figure}

In our AR(4) model, the coefficients of $\Delta \mathit{EXP}_{t-4}$ are
statically significant at the $1\%$ significance level because their
$p$-value is less than $0.01$ and the $t$-statistic exceeds the
critical value. The coefficient of $\Delta \mathit{EXP}_{t-3}$ is statically
significant at the $10\%$ significance. The constant and the other
coefficients are not statically significant. Even when the information
criteria are very low, this is not a good model because $R^2$ and
$\mathit{Adjusted}\ R^2$ are relatively small. So this AR(4) model turns out to
be not very useful to predict the growth rate in exports. Figure
\ref{deltaexp} shows that the frequency in this case is annual;
moreover, an increase in $\Delta \mathit{EXP}$ in one quarter is associated
with a decrease in the next one. In this case, the results of ADF test
allow us to reject the null hypothesis that rate of growth in export
has a unit autoregressive root both with the alternative hypothesis of
stationarity and of stationarity around a deterministic linear trend.
It follows that the QLR statistic is 5.02, which occurs in 2009:I, and
hence the hypothesis that the coefficients are stable is rejected at
the 1\% significance level. Again, the results of the software GRETL
confirm that the crisis of recent years has greatly affected the
exports from Verona. Consequently, by the results obtained we have
that the forecast of $\Delta \mathit{EXP}$ for 2014, given in the table
\begin{center}
{\footnotesize
\begin{tabular}{ccc}
\hline
$\mathit{Quarter}$ & $\mathit{Forecast}$ & $\mathit{Error}$
\\
\hline
2014:I & --4.86\% & 0.052736 \\

2014:II & 5.11\% & 0.053525 \\

2014:III & --1.58\% & 0.053961 \\

2014:IV & 6.16\% & 0.055304 \\
\hline
\end{tabular}}\vspace*{3pt}
\end{center}
and also sketched in Fig.~\ref{deltaexprev}, is not very accurate, and
the predictions do not perceive the lower peaks of the variable, which
is confirmed by the low value of $R^2$.
\subsection{IMP} \label{IMP}
\begin{figure}[t]
\includegraphics{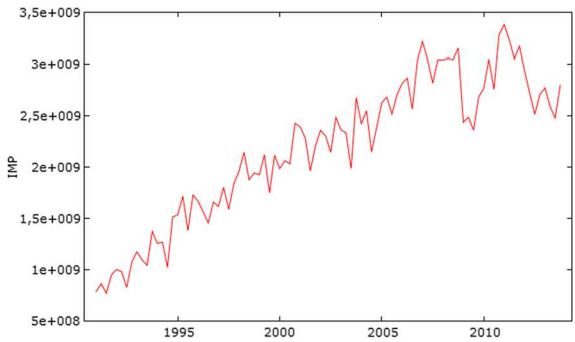}
\caption{Import of Verona}\label{imp}
\end{figure}

We now turn to the empirical problem to predict Verona import by
analyzing its historical series. We present an autoregressive model
that uses the history of Verona import to forecast its future. We use
92 observations of variable import, quarterly data from 1991 to 2013
expressed in Euros. Figure \ref{imp} shows the time series.

Looking at Fig.~\ref{imp}, we can see that Verona import shows
relatively smooth growth, although this decreases during the years
2008--2011; the curve is very similar to the time series of export, and
hence it is reasonable to deduce that decline in import is likely
caused by economic crisis broken out in Italy in those years. Although
the curve may seem apparently growing, periodic trends appear during
years under consideration. This curve has an annual periodicity.
Looking at a minimum of the curve, exactly one year later, another
minimum exists. The mean and standard deviation on a quarterly basis
for IMP are $ \mathit{Mean}= 2\,177\,300\,000 \  \mbox{\euro} $ and $ \mathit{Standard
Deviation} = 697\,420\,000 \  \mbox{\euro}$, whereas the annual mean export
is $ 2\,177\,300\,000 \times4 =\break 8\,709\,200\,000 \  \mbox{\euro}$. The first five
IMP autocorrelation values are $ \rho_1 = 0.9424$, $\rho_2 = \xch{\!0.9280}{0,9280}$,
$\rho_3 = 0.9060$, $\rho_4 = 0.9260$, $\rho_5 = 0.8750 $. These entries
show that inflation is strongly positively autocorrelated; in fact, the
first autocorrelation is 0.94. The autocorrelation remains large even
at the lag of four quarters. This means that an increase in import in
one quarter tends to be associated with an increase in the next
quarter. Autocorrelation, as expected, starts to decrease starting from
the lag of five quarters. As with the variable EXP, we estimated the AR
order of an autoregression in IMP using both the AIC and BIC
information criteria, finally obtaining that the optimal lag length is~4.

{\footnotesize
\begin{center}
\vspace{1em}
\begin{tabular}{lr@{.}lr@{.}lr@{.}lr@{.}l}
&
\multicolumn{2}{c}{Coefficient} &
\multicolumn{2}{c}{Standard Error} &
\multicolumn{2}{c}{$t$-Statistic} &
\multicolumn{2}{c}{$p$-Value} \\[1ex]
const &
1&90005\textrm{e$+$008} &
6&15140\textrm{e$+$007} &
3&0888 &
0&0027 \\
$\mathit{IMP}_{t-1}$ &
0&499665 &
0&0997006 &
5&0117 &
0&0000 \\
$\mathit{IMP}_{t-2}$ &
0&155637 &
0&0746261 &
2&0856 &
0&0401 \\
$\mathit{IMP}_{t-3}$ &
$-$0&154911 &
0&0881396 &
$-$1&7576 &
0&0825 \\
$\mathit{IMP}_{t-4}$ &
0&434062 &
0&0827892 &
5&2430 &
0&0000 \\
\end{tabular}

\vspace{1ex}
\begin{tabular}{lrlr}
SER & 198\,000\,000 \\
$R^2$ & 0.911613 & Adjusted $R^2$ & 0.907354 \\
AIC & 3616.812 & BIC & 3629.198\\
\end{tabular}
\end{center}
} Therefore, we have
\begin{align}
\label{AR(4)} %
\mathit{IMP} &=  190\,005\,000 + 0.499665 \mathit{IMP}_{t-1} + 0.155637 \mathit{IMP}_{t-2}\nonumber\\
& \quad  - 0.154911 \mathit{IMP}_{t-3} + 0.434062 \mathit{IMP}_{t-4}.
\end{align}

We check now if the model has a trend. The null hypothesis that Verona
import has a stochastic trend can be tested against the alternative
that it is stationary by performing the ADF test for a unit
autoregressive root. The ADF regression with four delays of IMP gives
\begin{align}
\label{ADFeq2} %
 \widehat{\Delta \mathit{IMP}}_t &= 190005000 + \delta \mathit{IMP}_{t-1} + \gamma_1 \Delta \mathit{IMP}_{t-1} + \gamma_2 \mathit{IMP}_{t-2}\nonumber\\
& \quad + \gamma_3 \Delta \mathit{IMP}_{t-3} + \gamma _4 \Delta \mathit{IMP}_{t-4}.
\end{align}
The ADF $t$-statistic is the $t$-statistic testing the hypothesis that
the coefficient on $\mathit{IMP}_{t-1}$ is zero, and it turns to be $t = -1.78$.
From Table \ref{ADF}, the 5\% critical value is $-2.86$. Because the
ADF statistic of \xch{$-1.78$}{$-1,78$} is less negative than $-2.86$, the test does
not reject the null hypothesis at the 5\% significance level. We
therefore cannot reject the null hypothesis that import has a unit
autoregressive root, that is, that import contains a stochastic trend,
against the alternative that it is stationary. If the alternative
hypothesis is that $Y_t$ is stationary around a deterministic linear
trend, then the ADF $t$-statistic results in \xch{$t = -2.6$}{$t = -2,6$}, which is less
negative than \xch{$-3.41$}{$-3,41$}. So, in this case, we also cannot reject the
null hypothesis that export has a unit autoregressive root.

\begin{figure}[t]
\includegraphics[scale=1.05]{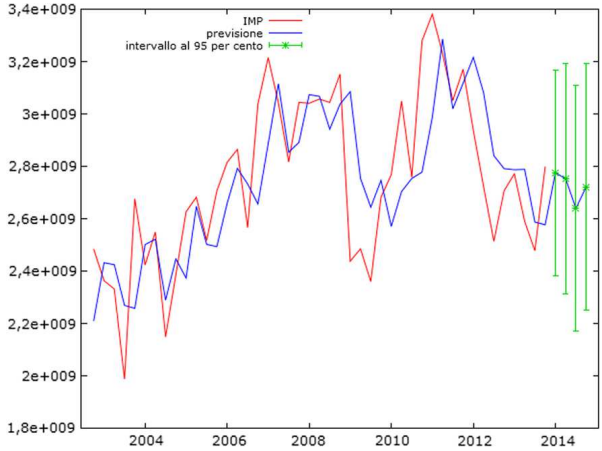}
\caption{Forecasts of Verona import}\label{imprev}\vspace*{3pt}
\end{figure}

We proceed with a QLR test, which provides a way to check whether the
import curve has been stable during the years sparing from 1993 to
2010. The Chow F-statistic tests the hypothesis that the intercept and
the coefficients at $\mathit{IMP}_{t-1} , \ldots,\break \mathit{IMP}_{t-4}$ in
Eq.~\eqref{AR(4)} are constant against the alternative that they break
at a given date for breaks in the central 70\% of the sample. Each
F-statistic tests five restrictions. The largest of these F-statistics
is 10.26, which occurs in 1995:III; the critical values for the
five-restriction model at different levels of significance are given
\xch{in Table~\ref{QLR5}}{in the following table}.
\begin{table}[b]
\caption{Critical values of QLR statistic with 15\% truncation}\label{QLR5}
\begin{tabular}{cccc}
\hline
$\mathit{Number \ of \ restrictions}$ & ${10 \ \%}$ &
${5 \ \%}$ & ${1 \ \%}$ \\
\hline
5 & 3.26 & 3.66 & 4.53 \\
\hline
\end{tabular}
\end{table}
These values indicate that the hypothesis that the coefficients are
stable is rejected at the 1\% significance level. Thus, there is an
evidence that at least one of these five coefficients changed over the
sample; namely, we have a structural break, which might be caused by
the devaluation that the Lira currency experienced during the period
1992--1995. According to the previous analysis, the predictions of
import of Verona for the year 2014 are as follows:\vspace*{3pt}
\begin{center}
{\footnotesize
\begin{tabular}{ccc}
\hline
$\mathit{Quarter}$ & $\mathit{Forecast}$ & $\mathit{Error}$
\\
\hline
2014:I & 2\,775\,360\,000 & 197\,957\,000 \\

2014:II & 2\,752\,530\,000 & 197\,957\,000 \\

2014:III & 2\,639\,510\,000 & 235\,388\,000 \\

2014:IV & 2\,721\,670\,000 & 236\,693\,000 \\
\hline
\end{tabular}}
\end{center}
They result in a slight increase for the next year, as shown by \xch{Fig.}{the graph}~\ref{imprev}.

\vspace*{3pt}\subsection{$\Delta \mathit{IMP}$}\vspace*{3pt}
The fourth variable of interest is represented by the logarithm of the
ratio between consecutive values of IMP, that is,\vspace*{3pt}
\[
\Delta \mathit{IMP}_t = \ln(\mathit{IMP}_t) - \ln(\mathit{IMP}_{t-1}) =
\ln \biggl( \frac
{\mathit{IMP}_t}{\mathit{IMP}_{t-1}} \biggr).
\]
The first six autocorrelations values of $\Delta \mathit{IMP}$ \xch{are presented in Table~\ref{auto}.}{are as follows:}
\begin{table}[t]
\caption{Autocorrelations of ${\Delta \mathit{IMP}}$}\label{auto}
\begin{tabular}{ccccccc}
\hline
$\boldsymbol{j}$ & 1 & 2 & 3 & 4 & 5 & 6\\
\hline
$\boldsymbol{\rho_j}$ & \xch{--0.4240}{-0,4240} & \xch{0.0631}{0,0631} & \xch{--0.3910}{-0,3910} & \xch{0.6721}{0,6721} & \xch{--0.3844}{-0,3844} &
\xch{0.0743}{0,0743} \\
\hline
\end{tabular}
\end{table}

\begin{figure}[t]
\includegraphics{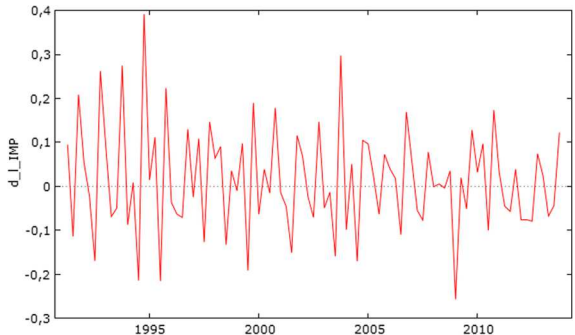}
\caption{Rate of growth in imports}\label{deltaimp}
\end{figure}

In the case of the growth rate of export, the negative
autocorrelation of the change of import means that, on average, an
increase in import in one quarter is associated with a decrease in the
next one. From the fifth lag, autocorrelation starts to be less
significant. So, it can be easily seen from Fig.~\ref{deltaimp} and
the autocorrelations in Table \ref{auto} that the right estimate of the
lag length is 4. The consequent AR(4) model reads as follows:\vspace*{3pt}
\begin{align}
\label{AR(4)2} %
\Delta \mathit{IMP} &= \xch{0.0128189}{0,0128189} - \xch{0.173627}{0,173627} \Delta \mathit{IMP}_{t-1} + \xch{0.0996175}{0,0996175} \Delta \mathit{IMP}_{t-2}\nonumber\\
& \quad - \xch{0.189882}{0,189882} \Delta \mathit{IMP}_{t-3} + \xch{0.416414}{0,416414} \Delta \mathit{IMP}_{t-4} \,,
\end{align}
and the following Fig.~\ref{deltaimp} shows the time series of $\Delta
\mathit{IMP}$, and we can see how an increase in import in one quarter is
associated with a decrease in the next one.

\begin{center}

\vspace{1em} {\footnotesize
\begin{tabular}{lr@{.}lr@{.}lr@{\xch{.}{,}}lr@{.}l}
&
\multicolumn{2}{c}{Coefficient} &
\multicolumn{2}{c}{Standard Error} &
\multicolumn{2}{c}{$t$-Statistic} &
\multicolumn{2}{c}{p-Value} \\[1ex]
const                       & 0    & 0161472 & 0 & 0110277 & 1    & 4642 & 0 & 1470 \\
$\Delta \mathit{IMP}_{t-1}$ & $-$0 & 326437  & 0 & 0950214 & $-$3 & 4354 & 0 & 0009 \\
$\Delta \mathit{IMP}_{t-2}$ & $-$0 & 224760  & 0 & 0878146 & $-$2 & 5595 & 0 & 0123 \\
$\Delta \mathit{IMP}_{t-3}$ & $-$0 & 280232  & 0 & 0960526 & $-$2 & 9175 & 0 & 0046 \\
$\Delta \mathit{IMP}_{t-4}$ & 0    & 431620  & 0 & 0894247 & 4    & 8266 & 0 & 0000 \\[4.5pt]
\end{tabular}
\vspace{1ex}
\begin{tabular}{lrlr}
SER & 0.083791 \\
$R^2$ & 0.531621 & Adjusted $R^2$ & 0.508773 \\
AIC & --179.6755 & BIC & --167.3459\\
\end{tabular}
}
\end{center}

\begin{figure}[t]
\includegraphics{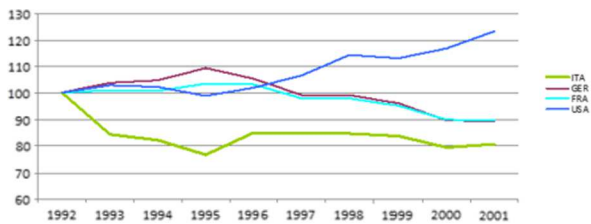}
\caption{Evolution of the Real Exchange Rate (index numbers: 1992 $=$
100) (\emph{Source: FMI})}\label{cambio}
\end{figure}

The QLR statistic for AR(4) model in Eq.~\eqref{AR(4)2} is 22.58,
which occurs in 1995:II. This value indicates that the hypothesis that
the coefficients are stable is rejected at the 1\% significance level.
As for imports, we can associate this structural break to the last
crisis of Lira occurred in that period. We observe the dynamics of the
real effective exchange rate in Fig.~\ref{cambio}.

\begin{figure}[t]
\includegraphics{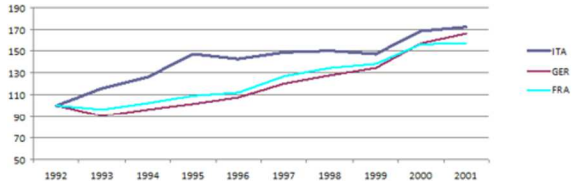}
\caption{Growth of Exports of Goods and Services (index numbers: 1992 $=$
100; correct values with the GDP deflator) (\emph{Source: World Bank
data})}\label{export}
\end{figure}

As shown in Fig.~\ref{export}, the devaluation of the Lira has produced
some benefits for the growth of Italian exports (goods and services),
especially looking at analogous economical data for Germany and France.

\begin{figure}[t]
\includegraphics{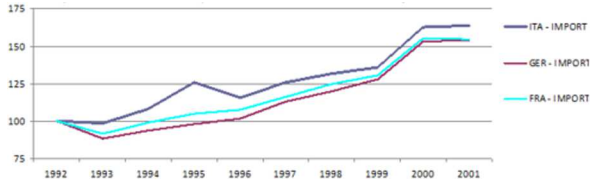}
\caption{Growth of Imports of Goods and Services (index numbers: 1992 $=$
100; correct values with the GDP deflator) (\emph{Source: World Bank
data})}\label{import}
\end{figure}

As shown in Fig.~\ref{import}, the devaluation of the Lira did not stop
the value of imports, but you can still easily perceive the rupture of
1995.

\subsection{Active Enterprises}
\begin{figure}[t]
\includegraphics{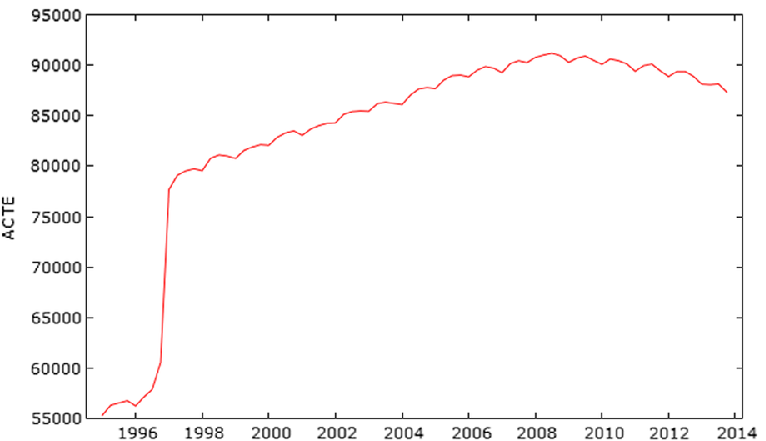}
\caption{$\mathit{ACTE}_t$}\label{acte}
\end{figure}

We would like also to briefly analyze the variable ``Active
Enterprises'' ($\mathit{ACTE}_t$), namely the time series with quarterly data
from 1995 to 2013, where each observation is the number of firms
operating in a given quarter in the province of Verona. With the
software GRETL we obtain the AR(4) model
\begin{align}
\label{AR(4)3} %
\mathit{ACTE} &=  9535.97 + 1.02210 \mathit{ACTE}_{t-1} - 0.173385 \mathit{ACTE}_{t-2}\nonumber\\
& \quad + 0.0152586 \mathit{ACTE}_{t-3} + 0.0280194 \mathit{ACTE}_{t-4} \,.
\end{align}
The $\mathit{Adjusted}\ R^2$ of this regression is 0.94, and the QLR statistic is
37.52, which occurs in 2011:I. This value indicates that the
hypothesis that the coefficients are stable is rejected at the 1\%
significance level. Also, for the variable $\mathit{ACTE}_t$, we can conclude
that the number of active businesses were affected by the crisis of
those years. However, the ADF $t$-statistic for this variable does not
reject the null hypothesis, so we cannot reject the fact that the time
series of the numbers of active enterprises has a unit autoregressive
root, that is, that $\mathit{ACTE}_t$ contains a stochastic trend, against the
alternative that it is stationary. From Fig.~\ref{acte} we can see that
the curve has a quite regular annual pattern and that active
enterprises tend to decline in the first quarter of each year and then
return generally to grow. It is worth to mention the drastic rise of
the curve during the first period of the time interval under
consideration. Such an increase has been caused by a particular type of
bureaucratic constraints, namely by a sort of forced registration
imposed to a rather large set of farms companies previously not obliged
to be part of the companies register. Such a norm has been introduced
in two steps, first by a simple communication (1993), and later in the
form of legal disclosure (2001).

\section{VAR models analysis of Verona data}

In this section, we apply the theory developed in the fourth chapter to
analyze the set of Verona import and export time series. Therefore, we
consider a VAR model for exports ($\mathit{EXP}_t$), imports ($\mathit{IMP}_t$), and
active companies ($\mathit{ACTE}_t$) in Verona, and each of such variables is
characterized by time series constituted by quarterly data from 1995 to
2013.

\subsection{First model: stationary variables}
\begin{figure}[t]
\includegraphics{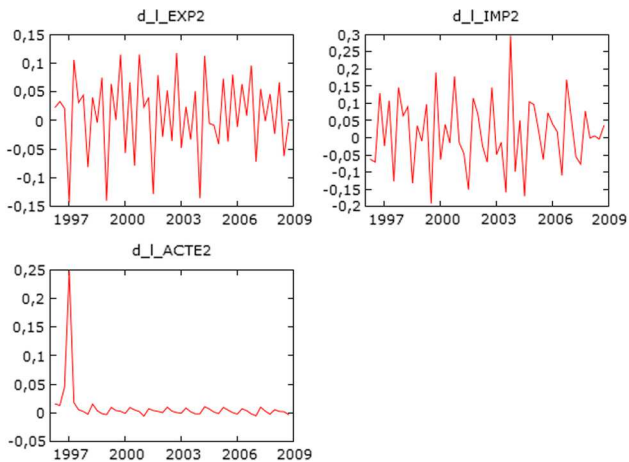}
\caption{Multiple graph for ${\Delta \mathit{EXP}_t}$,
${\Delta \mathit{EXP}_t}$ and ${\Delta
\mathit{ACTE}_t}$}\label{multiple}\vspace*{-6pt}
\end{figure}

As we saw in Chapter \ref{chapter3}, the import end export of Verona are
subject to a stochastic trend, so that it is appropriate to transform
it by computing its logarithmic first differences in order to obtain
stationary variables. Figure \ref{multiple} shows a multiple graph for
the time series of ${\Delta \mathit{EXP}_t}$, ${\Delta \mathit{IMP}_t}$, and ${\Delta
\mathit{ACTE}_t}$.

The VAR for $\Delta \mathit{EXP}_t$, $\Delta \mathit{IMP}_t$, and $\Delta \mathit{ACTE}_t$
consists of three equations, each of which is characterized by a
dependent variable, namely by $\Delta \mathit{EXP}_t$, $\Delta \mathit{IMP}_t$, and
$\Delta \mathit{ACTE}_t$, respectively. Because of the apparent breaks in
considered time series for the years 1995 and 2010, the VAR is
estimated using data from 1996:I to 2008:IV. The number of lags of this
model are obtained through information criteria BIC and AIC using the
software GRETL, which gives the results in Table \ref{VARll}, where the
asterisks indicate the best (or minimized) of the respective
information criteria.\vadjust{\eject}

\begin{table}[t]
\caption{VAR lag lengths}\label{VARll}
\begin{tabular}{cll}
\hline
$p$ & $\mathit{AIC(p)}$ & $\mathit{BIC(p)}$ \\
\hline
1 & --13.610968 & --13.119471 \\
2 & --14.685333 & --13.825212* \\
3 & --14.572491 & --13.343746 \\
4 & --14.747567 & --13.150199 \\
5 & --14.974180 & --13.008189 \\
6 & --15.160238* & --12.825624 \\
7 & --15.048342 & --12.345105 \\
8 & --15.047682 & --11.975822 \\
\hline
\end{tabular}
\end{table}

The smallest AIC has been obtained considering six lags; indeed, the
BIC estimation of the lag length is $\hat{p} = 2$. We decide to choose
two delays because, for $\hat{p} = 6$, we have a VAR with three
variables and six lags, so we will have 19 coefficients (eight lags
with three variables each, plus the intercept) in each of the three
equations, with a total of 57 coefficients, and we saw in
\cite[Sect.~4.2]{rif9} that estimation of all these coefficients
increases the amount of the forecast estimation error, resulting in a
deterioration of the accuracy of the forecast itself. We also prefer
consider the BIC estimation for its consistency; however, the AIC
overestimate $p$ (see \cite[Sect.~2.2]{rif9}. Estimating the VAR model
with GRETL produces the following results:
\begin{align}
\label{deltaVAR} %
\Delta \mathit{EXP}_t &= 0.0014 - 0.44 \Delta \mathit{EXP}_{t-1} - 0.14 \Delta \mathit{EXP}_{t-2} - 0.19 \Delta \mathit{IMP}_{t-1}\nonumber\\
&\quad{}+ 0.21 \Delta \mathit{IMP}_{t-2} - 0.15 \Delta \mathit{ACTE}_{t-1} + 0.35 \Delta \mathit{ACTE}_{t-2},\nonumber\\
\Delta \mathit{IMP}_t &= 0.0222 - 0.5 \Delta \mathit{EXP}_{t-1} + 0.57 \Delta \mathit{EXP}_{t-2} - 0.38 \Delta \mathit{IMP}_{t-1}\nonumber\\
&\quad{}- 0.46 \Delta \mathit{IMP}_{t-2} + 0.09 \Delta \mathit{ACTE}_{t-1} + 0.2 \Delta \mathit{ACTE}_{t-2},\nonumber\\
\Delta \mathit{ACTE}_t &= 0.0043 + 0.02 \Delta \mathit{EXP}_{t-1} + 0.12 \Delta \mathit{EXP}_{t-2} + 0.07 \Delta \mathit{IMP}_{t-1}\nonumber\\
&\quad{}- 0.02 \Delta \mathit{IMP}_{t-2} + 0.23 \Delta \mathit{ACTE}_{t-1} + 0.02 \Delta \mathit{ACTE}_{t-2}.
\end{align}

In the first equation ($\Delta \mathit{EXP}_t$) of VAR system
\eqref{deltaVAR}, we have the coefficients of $\Delta \mathit{EXP}_{t-1}$,
$\Delta \mathit{IMP}_{t-2}$, and $\Delta \mathit{ACTE}_{t-2}$, which are statically
significant at the $1\%$ significance level because their $p$-value is
less than $0.01$ and the $t$-statistic exceeds the critical value. The
constant and the coefficients of $\Delta \mathit{IMP}_{t-1}$, however, are
statically significant at the $5\%$ significance, and the other
coefficients are not statically significant. The $\mathit{Adjusted}\ R^2$ is
0.53. In the second equation ($\Delta \mathit{IMP}_t$) of VAR system
\eqref{deltaVAR}, we have the coefficients of $\Delta \mathit{EXP}_{t-1}$,
$\Delta \mathit{EXP}_{t-2}$, $\Delta \mathit{IMP}_{t-1}$, and $\Delta \mathit{IMP}_{t-2}$, which
are statically significant at the $1\%$ significance level. The
constant, however, is statically significant at the $10\%$
significance, and the other coefficients are not statically
significant. The $\mathit{Adjusted}\ R^2$ is 0.45. In the last equation of
\eqref{deltaVAR}, we have only the constant statically significant, at
the $5\%$ level. The $\mathit{Adjusted}\ R^2$ is \xch{$-0.04$}{--0,04}. These VAR equations can
be used to perform Granger causality tests. The results of this test
for the first equation of \eqref{deltaVAR} are as follows:
\begin{table}[h!]
\begin{tabular}{ccc}
\hline
Variable & Test F & $p$-Value \\
\hline
$\Delta \mathit{IMP}_t$ & 12.464 & 0.0001 \\
$\Delta \mathit{ACTE}_t$ & 8.2240 & 0.0010 \\
\hline
\end{tabular}
\end{table}
\\The F-statistic testing the null hypothesis that the coefficients of
${\Delta \mathit{IMP}_{t-1}}$ and\break ${\Delta \mathit{IMP}_{t-2}}$ are zero in the first
equation is 12.46 with $p$-value 0.0001, which is less than 0.01. Thus,
the null hypothesis is rejected at the level of 1\%, so we can conclude
that the growth rate in Verona import is a useful predictor for the
growth rate in export, namely ${\Delta \mathit{IMP}_t}$ Granger-causes ${\Delta
\mathit{EXP}_t}$. Also, ${\Delta \mathit{ACTE}_t}$ Granger-causes the change in export at
the 1\% significance level. The results for the second equation of
\eqref{deltaVAR} are
as follows:\vspace*{-6pt}
\begin{table}[h]
\begin{center}
\begin{tabular}{ccc}
\hline
Variable & Test F & $p$-Value \\
\hline
$\Delta \mathit{EXP}_t$ & \xch{22.766}{22,766} & \xch{0.0000}{0,0000} \\
$\Delta \mathit{ACTE}_t$ & \xch{1.5894}{1,5894} & \xch{0.2161}{0,2161} \\
\hline
\end{tabular}\vspace*{-6pt}
\end{center}
\end{table}
\\For the ${\Delta \mathit{IMP}_t}$ equation, we can also conclude that the
growth rate in Verona export is a useful predictor for the growth rate
in import, but the change in the number of active enterprises is not.
The results for the last equation of \eqref{deltaVAR} are as follows:
\begin{table}[h]
\begin{center}
\begin{tabular}{ccc}
\hline
Variable & Test F & $p$-Value \\
\hline
$\Delta \mathit{EXP}_t$ & \xch{1.0897}{1,0897} & \xch{0.3456}{0,3456} \\
$\Delta \mathit{EXP}_t$ & \xch{1.6413}{1,6413} & \xch{0.2059}{0,2059} \\
\hline
\end{tabular}\vspace*{-6pt}
\end{center}
\end{table}
\\The F-statistic testing the null hypothesis that the coefficients of
${\Delta \mathit{EXP}_{t-1}}$ and\break ${\Delta \mathit{EXP}_{t-2}}$ are zero in the first
equation is 1.09 with $p$-value 0.34, which is greater than \xch{0.10}{0,10}. Thus,
the null hypothesis is not rejected, so we can conclude that the growth
rate in Verona import is not a useful predictor for the growth rate in
active enterprises, namely, ${\Delta \mathit{IMP}_t}$ does not Granger-cause
${\Delta \mathit{ACTE}_t}$. The F-statistic testing the hypothesis that the
coefficients of the two lags of ${\Delta \mathit{EXP}_t}$ are zero is 1.64 with
$p$-value of 0.2; thus, ${\Delta \mathit{EXP}_t}$ also does not Granger-cause
${\Delta \mathit{ACTE}_t}$ at the 10\% significance level. Forecasts of the
three variables in system \eqref{deltaVAR} are obtained exactly as
discussed in the univariate time series models, but in this case, the
forecast of ${\Delta \mathit{EXP}_t}$, we also consider past values of ${\Delta
\mathit{IMP}_t}$ and ${\Delta \mathit{ACTE}_t}$.

{\scriptsize
\begin{longtable}{%
r
r@{.}l
r@{.}l
r@{.}l
r@{.}l
r@{.}l}
\caption{Forecasts of ${\Delta \mathit{EXP}_t}$}\\
\hline
Quarter & \multicolumn{2}{c}{${\Delta \mathit{EXP}_t}$} &
\multicolumn{2}{c}{Forecast}
& \multicolumn{2}{c}{Error}
& \multicolumn{4}{c}{95\% Confidence Interval} \\
\hline
2009:1 & \multicolumn{2}{c}{--0.02329} & 0&018816 & 0&044544 &
$-$0&071077 & 0&108709 \\
2009:2 & \multicolumn{2}{c}{0.06674} & 0&017759 & 0&052249 &
$-$0&087683 & 0&123202 \\
2009:3 & \multicolumn{2}{c}{--0.06221} & $-$0&002095 & 0&060086 &
$-$0&123354 & 0&119164 \\
2009:4 & \multicolumn{2}{c}{--0.003635} & 0&006493 & 0&063164 &
$-$0&120976 & 0&133963 \\
2010:1 & \multicolumn{2}{c}{--0.1911938} & 0&016435 & 0&065429 &
$-$0&115605 & 0&148475 \\
2010:2 & \multicolumn{2}{c}{0.0002207} & 0&013870 & 0&066425 &
$-$0&120182 & 0&147922 \\
2010:3 & \multicolumn{2}{c}{--0.03853} & 0&004754 & 0&067609 &
$-$0&131686 & 0&141194 \\
2010:4 & \multicolumn{2}{c}{0.08106} & 0&009609 & 0&068220 &
$-$0&128063 & 0&147282 \\
2011:1 & \multicolumn{2}{c}{--0.002692} & 0&013526 & 0&068644 &
$-$0&125004 & 0&152055 \\
2011:2 & \multicolumn{2}{c}{0.1127259} & 0&011529 & 0&068840 &
$-$0&127397 & 0&150454 \\
2011:3 & \multicolumn{2}{c}{--0.02047} & 0&007798 & 0&069059 &
$-$0&131568 & 0&147164 \\
2011:4 & \multicolumn{2}{c}{0.08649} & 0&010466 & 0&069186 &
$-$0&129156 & 0&150088 \\
2012:1 & \multicolumn{2}{c}{--0.04747} & 0&011927 & 0&069269 &
$-$0&127863 & 0&151716 \\
2012:2 & \multicolumn{2}{c}{0.06716} & 0&010711 & 0&069309 &
$-$0&129160 & 0&150583 \\
2012:3 & \multicolumn{2}{c}{--0.003477} & 0&009260 & 0&069350 &
$-$0&130694 & 0&149213 \\
2012:4 & \multicolumn{2}{c}{0.07761} & 0&010650 & 0&069377 &
$-$0&129358 & 0&150658 \\
2013:1 & \multicolumn{2}{c}{--0.07186} & 0&011142 & 0&069393 &
$-$0&128898 & 0&151182 \\
2013:2 & \multicolumn{2}{c}{0.05621} & 0&010477 & 0&069401 &
$-$0&129580 & 0&150535 \\
2013:3 & \multicolumn{2}{c}{--0.04800} & 0&009946 & 0&069409 &
$-$0&130127 & 0&150019 \\
2013:4 & \multicolumn{2}{c}{0.06604} & 0&010640 & 0&069415 &
$-$0&129445 & 0&150724 \\
2014:1 & \multicolumn{2}{c}{--0.09138} & 0&010775 & 0&069418 &
$-$0&129315 & 0&150866 \\
2014:2 & \multicolumn{2}{c}{0.05304} & 0&010436 & 0&069420 &
$-$0&129658 & 0&150530 \\
2014:3 & \multicolumn{2}{c}{--0.001114} & 0&010259 & 0&069421 &
$-$0&129838 & 0&150356 \\
2014:4 & \multicolumn{2}{c}{0.07616} & 0&010592 & 0&069422 &
$-$0&129507 & 0&150692\\
\hline
\end{longtable}}

\begin{figure}[t]
\includegraphics{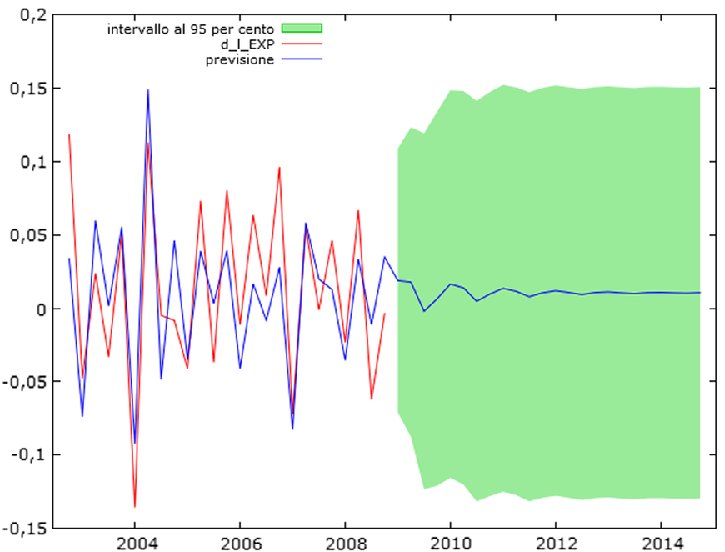}
\caption{Forecast for \xch{${\Delta \mathit{EXP}_t}$ (color online)}{$\boldsymbol{\Delta \mathit{EXP}_t}$}}\label{1}
\end{figure}

\begin{figure}[t]
\includegraphics{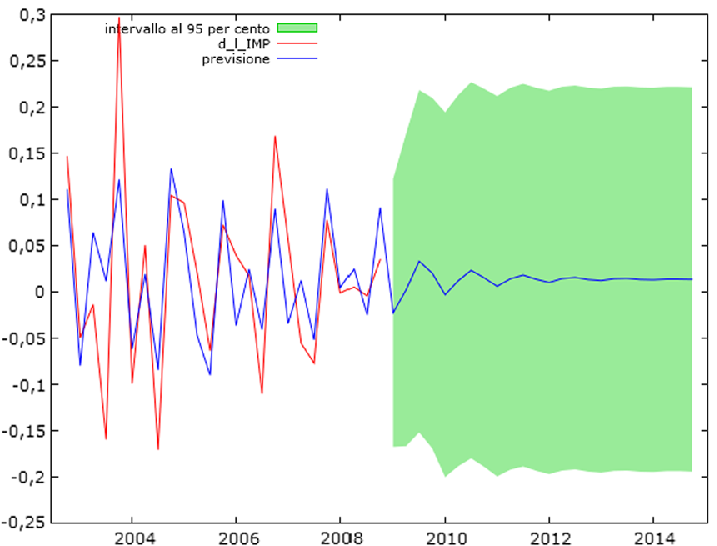}
\caption{Forecast for \xch{${\Delta \mathit{IMP}_t}$ (color online)}{$\boldsymbol{\Delta \mathit{IMP}_t}$}}\label{2}
\end{figure}

\begin{figure}[t]
\includegraphics{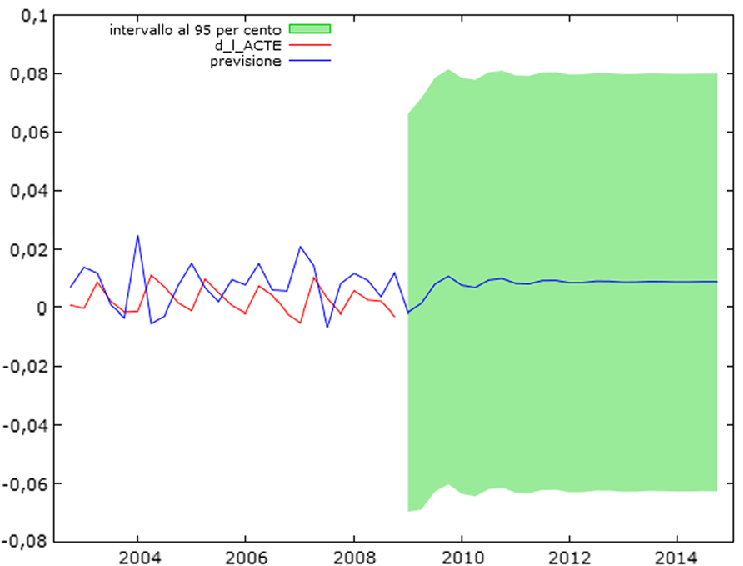}
\caption{Forecast for \xch{${\Delta \mathit{ACTE}_t}$ (color online)}{$\boldsymbol{\Delta \mathit{ACTE}_t}$}}\label{3}
\end{figure}

By means of the forecasts from 2009 to 2013, we can establish a
comparison with the real data, noting that the predictions with this
VAR model are not very reliable since the error is quite high and it
increases in recent years. The lack of accuracy was confirmed
previously by low values of the $\mathit{Adjusted}\ R^2$. \xch{Figures}{Graphs} \ref{1},
\ref{2}, and \ref{3} show the real time series of the three variables
with a~red line and the prediction made with the estimated models with
a blue line. It can be seen from these graphs that the confidence
intervals (green area in the figures) are very high.

\subsection{Second model: nonstationary variables}
\begin{figure}[t]
\includegraphics{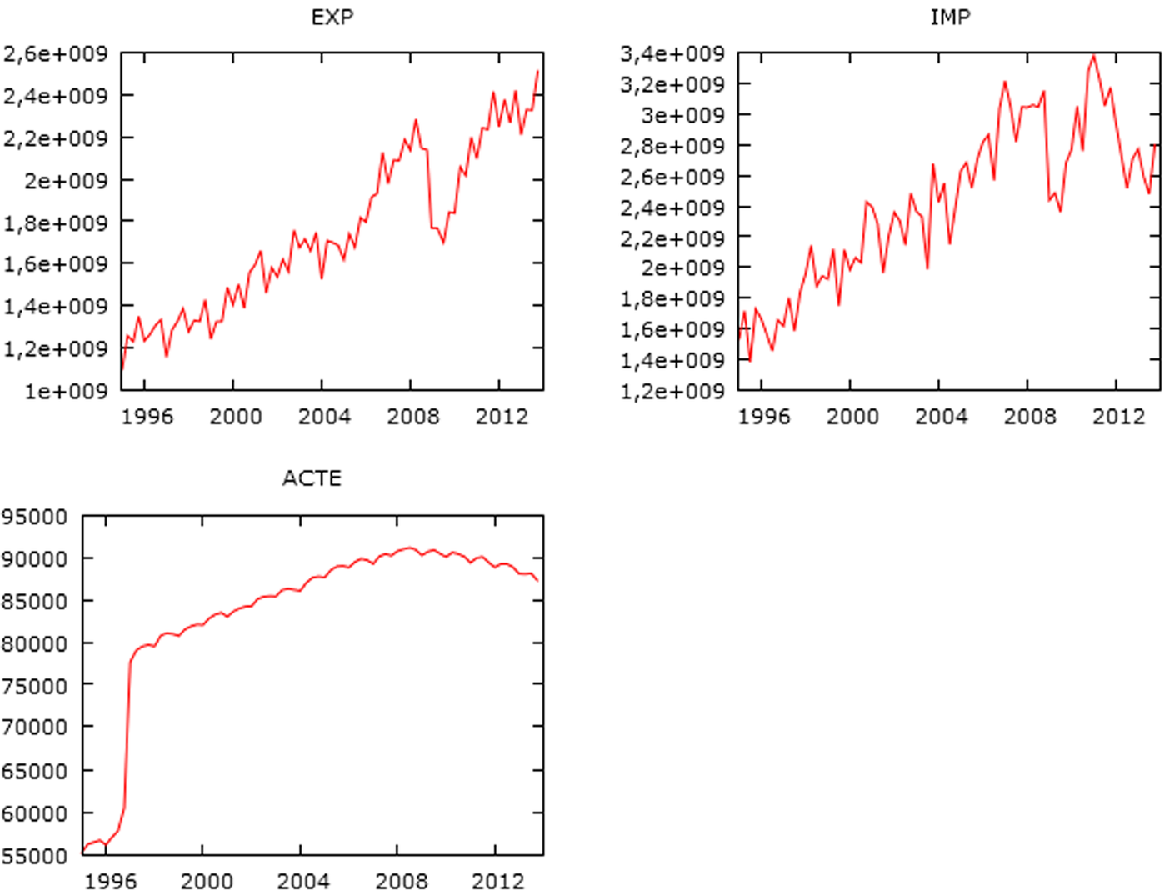}
\caption{Multiple Graph for ${\mathit{EXP}_t}$, ${\mathit{IMP}_t}$,
and ${\mathit{ACTE}_t}$}\label{multiple2}
\end{figure}

In this section, we analyze the three variable ($\mathit{EXP}_t$, $\mathit{IMP}_t$, and
$\mathit{ACTE}_t$), considering quarterly Verona data from 1995 to 2013. We
analyze these time series without avoiding structural breaks and
without considering the first differences, and we check if the analysis
produces different results with respect to the previous ones. Figure
\ref{multiple2} shows a multiple graph for the time series respectively
of $\mathit{EXP}_t$, $\mathit{IMP}_t$, and $\mathit{ACTE}_t$.

The GRETL lag length selection gives the results in Table \ref{VARll2};
then, according to the considerations made to determine the number of
delays for the model \eqref{deltaVAR}, we decide to choose three
delays, obtaining the following model:
\begin{table}[t]
\caption{VAR lag lengths}\label{VARll2}
\begin{tabular}{cll}
\hline
$p$ & $\mathit{AIC(p)}$ & $\mathit{BIC(p)}$ \\
\hline
1 & 98.133995 & 98.525673 \\
2 & 97.843826 & 98.529262 \\
3 & 97.432952 & 98.412147* \\
4 & 97.327951 & 98.600904 \\
5 & 97.204547 & 98.771258 \\
6 & 97.107478 & 98.967947 \\
7 & \xch{97.020628}{97,020628} & 99.174856 \\
8 & 97.007994* & 99.455981 \\
\hline
\end{tabular}
\end{table}
%
\begin{align}
\label{nodeltaVAR}
\mathit{EXP}_t &= -119\,893\,000 + \xch{0.79}{0,79} \mathit{EXP}_{t-1} + \xch{0.47}{0,47} \mathit{EXP}_{t-2} - \xch{0.31}{0,31} \mathit{EXP}_{t-3} \nonumber\\
& \quad - \xch{0.12}{0,12} \mathit{IMP}_{t-1} + \xch{0.20}{0,20} \mathit{IMP}_{t-2} - \xch{0.09}{0,09} \mathit{IMP}_{t-3} \nonumber\\
& \quad + \xch{4389.61}{4389,61} \mathit{ACTE}_{t-1} + \xch{4715.09}{4715,09} \mathit{ACTE}_{t-2} - \xch{6479.85}{6479,85} \mathit{ACTE}_{t-3}, \nonumber\\
\mathit{IMP}_t &= -313\,115\,000 + \xch{0.17}{0,17} \mathit{EXP}_{t-1} + \xch{1.11}{1,11} \mathit{EXP}_{t-2} - \xch{1.21}{1,21} \mathit{EXP}_{t-3} \nonumber\\
& \quad + \xch{0.52}{0,52} \mathit{IMP}_{t-1} - \xch{0.13}{0,13} \mathit{IMP}_{t-2} + \xch{0.28}{0,28} \mathit{IMP}_{t-3} \nonumber\\
& \quad + \xch{13\,719.5}{13719,5} \mathit{ACTE}_{t-1} - \xch{3215.16}{3215,16} \mathit{ACTE}_{t-2} + \xch{1103.38}{1103,38} \mathit{ACTE}_{t-3}, \nonumber\\
\mathit{ACTE}_t &= \xch{8526.18}{8526,18} - \xch{1.62}{1,62} \times10^{-6} \mathit{EXP}_{t-1} + \xch{1.87}{1,87} \times10^{-6} \mathit{EXP}_{t-2} \nonumber\\
& \quad - \xch{7059}{7,59} \times10^{-8} \mathit{EXP}_{t-3} + \xch{1.31}{1,31} \times10^{-6} \mathit{IMP}_{t-1} \nonumber\\
& \quad - \xch{4.89}{4,89} \times10^{-7} \mathit{IMP}_{t-2} - \xch{3.81}{3,81} \times10^{-7} \mathit{IMP}_{t-3} + \xch{1.04}{1,04} \mathit{ACTE}_{t-1} \nonumber\\
& \quad - \xch{0.17}{0,17} \mathit{ACTE}_{t-2} + \xch{0.02}{0,02} \mathit{ACTE}_{t-3}.
\end{align}
%
\!\!In the first equation ($\mathit{EXP}_t$) of VAR system \eqref{nodeltaVAR}, we
have the coefficients of $\mathit{EXP}_{t-1}$, $\mathit{EXP}_{t-2}$, $\mathit{IMP}_{t-2}$, and
$\mathit{ACTE}_{t-3}$, which are statically significant at the $1\%$ level
because their $p$-value is less than $0.01$ and the $t$-statistic
exceeds the critical value. However, the coefficients of $\mathit{EXP}_{t-3}$
and $\mathit{IMP}_{t-1}$ are statically significant at the $5\%$ level, and the
other coefficients are not statically significant. The coefficient of
$\mathit{IMP}_{t-3}$ is statically significant at the $10\%$ level, and the
others are not statically significant. The $\mathit{Adjusted}\ R^2$ is 0.93. In
the second equation ($\mathit{IMP}_t$) of VAR system \eqref{nodeltaVAR}, we have
the coefficients of $\mathit{EXP}_{t-2}$, $\mathit{EXP}_{t-3}$, $\mathit{IMP}_{t-1}$,
$\mathit{IMP}_{t-3}$, and $\mathit{ACTE}_{t-1}$, which are statically significant at the
$1\%$ significance level. The constant is statically significant at the
$5\%$~level, and the other coefficients are not statically significant.
The $\mathit{Adjusted}\ R^2$ is 0.85. In the last equation of \eqref{nodeltaVAR},
we have only $\mathit{EXP}_{t-1}$, $\mathit{ACTE}_{t-1}$, and
$\mathit{ACTE}_{t-2}$ statically significant respectively at the $5\%, 1\%$,
and $10\%$ levels, whereas the $\mathit{Adjusted}\ R^2$ is 0.94. If we perform
Granger causality tests, then we have that all $p$-values of the
F-statistic of the three equations are less than 0.01; only for the
third equation of \eqref{nodeltaVAR}, the Granger causality test for
the variable $\mathit{EXP}_t$ has the $p$-value 0.0852, and hence $\mathit{EXP}_t$
Granger-causes $\mathit{ACTE}_t$, but in this case, the null hypothesis is
rejected at the level of 10\%. Notice that the model \eqref{nodeltaVAR}
has high values of the $\mathit{Adjusted}\ R^2$, so it can be very useful to make
prediction of future values of the three variables. The forecasts for
$\mathit{EXP}_t$ concerning 2014 are given by the table
\begin{center}
{\footnotesize
\begin{tabular}{ccc}
\hline
$\mathit{Quarter}$ & $\mathit{Forecast}$ & $\mathit{Error}$
\\
\hline
2014:I & 2\,415\,830\,000 & 86\,744\,900 \\

2014:II & 2\,502\,280\,000 & 105\,860\,000 \\

2014:III & 2\,430\,160\,000 & 143\,193\,000 \\

2014:IV & 2\,488\,770\,000 & 158\,629\,000 \\
\hline
\end{tabular}}\vadjust{\eject}
\end{center}

\begin{figure}[t]
\includegraphics{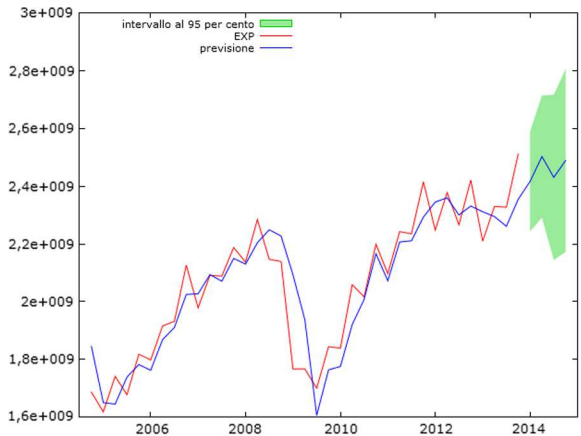}
\caption{Forecast for $\mathit{EXP}_t$}\label{4}
\end{figure}

Forecasts for ${\mathit{IMP}_t}$ are as follows:\vspace*{3pt}

\begin{center}
{\footnotesize
\begin{tabular}{ccc}
\hline
$\mathit{Quarter}$ & $\mathit{Forecast}$ & $\mathit{Error}$
\\
\hline
2014:I & 2\,764\,470\,000 & 174\,343\,000 \\

2014:II & 2\,870\,330\,000 & 200\,990\,000 \\

2014:III & 2\,712\,960\,000 & 237\,479\,000 \\

2014:IV & 2\,809\,610\,000 & 249\,185\,000 \\
\hline
\end{tabular}}\vspace*{3pt}
\end{center}
and prediction for $\mathit{ACTE}_t$ reads as follows:\vspace*{3pt}

\begin{center}
{\footnotesize
\begin{tabular}{ccc}
\hline
$\mathit{Quarter}$ & $\mathit{Forecast}$ & $\mathit{Error}$
\\
\hline
2014:I & \xch{87401.59}{87401,59} & \xch{1812.928}{1812,928} \\

2014:II & \xch{87988.15}{87988,15} & \xch{2634.586}{2634,586} \\

2014:III & \xch{88266.25}{88266,25} & \xch{3129.437}{3129,437} \\

2014:IV & \xch{88495.47}{88495,47} & \xch{3449.842}{3449,842} \\
\hline
\end{tabular}}\vspace*{3pt}
\end{center}

\begin{figure}[t]
\includegraphics{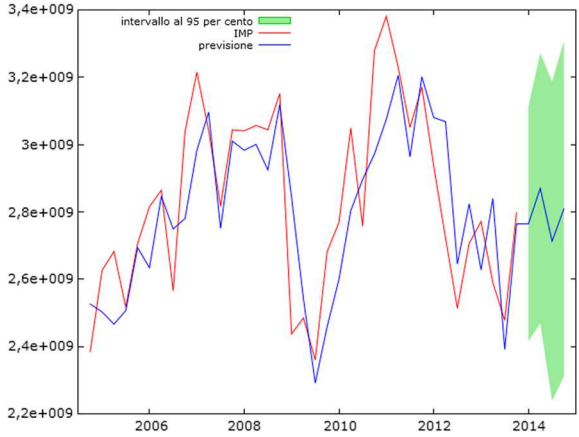}
\caption{Forecast for $\mathit{EXP}_t$}\label{5}\vspace*{-3pt}
\end{figure}

\begin{figure}[t]
\includegraphics{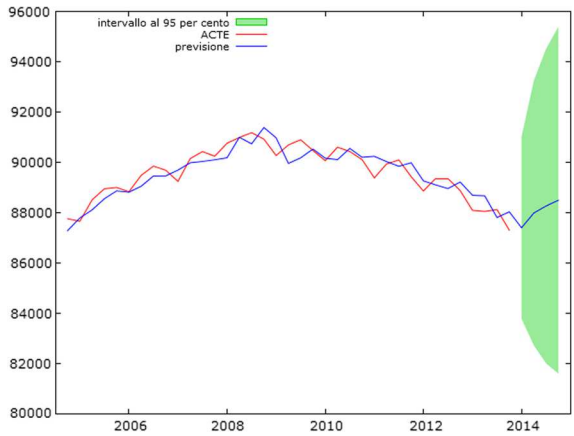}
\caption{Forecast for $\mathit{ACTE}_t$}\label{6}\vspace*{-3pt}
\end{figure}

Figures \ref{4}, \ref{5}, and \ref{6} show the time series of the three
variables and their forecasts. The area of confidence interval for
$\mathit{EXP}_t$ is rather small, which is confirmed by the value 0.93 of the
$\mathit{Adjusted}\ R^2$ of the first equation in system~\eqref{nodeltaVAR}. This
area is slightly wider for the second graph, and in Fig.~\ref{6} we
show the confidence interval for $\mathit{ACTE}_t$ becoming wider at each
quarter.

\subsection{No cointegration between $\mathit{EXP}_t$ and $\mathit{IMP}_t$}
We saw in \xch{Sections}{Subsections} \ref{EXP} and \ref{IMP} that the time series for
$\mathit{EXP}_t$ and $\mathit{IMP}_t$ are both integrated of order 1 (I(1)); hence, we
perform an EG-ADF test to verify if these two variables are
cointegrated. The cointegrating coefficient $\theta$ is estimated by
the OLS estimate of the regression $ \mathit{EXP}_t = \alpha+ \theta \mathit{IMP}_t +
z_t $; hence, we obtain $ \mathit{EXP}_t =197\,119\,000 + 0.641536 \mathit{IMP}_t + z_t$, so
that $\theta= 0.641536 $. Then we use a Dickey--Fuller test to test
for a unit root in $z_t = \mathit{EXP}_t - \theta \mathit{IMP}_t$. The statistic test
result is --2.77065, which is greater than --3.96 (see \cite[Table
1]{rif9} for critical values); therefore, we cannot refuse the null
hypothesis of a unit root for $z_t$, concluding that the series $\mathit{EXP}_t
- \theta \mathit{IMP}_t$ is~not stationary. Moreover, we have that the
variables $\mathit{EXP}_t$ and $\mathit{IMP}_t$ are not cointegrated.

%
%
%
%
%
%

\section{VAR model with Italian data}
In this section, we perform a comparison of the time series between
provincial and national data. Considering the same model of system
\eqref{nodeltaVAR}, but with data referring to Italy, we get a VAR(8)
model of the form
\begin{equation}
\label{VAR(8)} %
\!\!\!\!\!\begin{cases}\!\!
\begin{aligned}
\mathit{EXPn}_t &= \hat{\beta}_{10} + \hat{\beta}_{11} \mathit{EXPn}_{t-1} + \cdots+ \hat
{\beta}_{18} \mathit{EXPn}_{t-8} + \hat{\gamma}_{11} \mathit{IMPn}_{t-1} \\
&\quad  + \cdots+ \hat{\gamma}_{18} \mathit{IMPn}_{t-8} + \hat{\delta}_{11}
\mathit{ACTEn}_{t-1} + \cdots+ \hat{\delta}_{18} \mathit{ACTEn}_{t-8}, \\
\mathit{IMPn}_t &= \hat{\beta}_{20} + \hat{\beta}_{21} \mathit{EXPn}_{t-1} + \cdots+ \hat
{\beta}_{28} \mathit{EXPn}_{t-8} + \hat{\gamma}_{21} \mathit{IMPn}_{t-1} \\
&\quad + \cdots+ \hat{\gamma}_{28} \mathit{IMPn}_{t-8} + \hat{\delta
}_{21} \mathit{ACTEn}_{t-1} + \cdots+ \hat{\delta}_{28} \mathit{ACTEn}_{t-8}, \\
\mathit{ACTEn}_t &= \hat{\beta}_{30} + \hat{\beta}_{31} \mathit{EXPn}_{t-1} + \cdots+ \hat
{\beta}_{38} \mathit{EXPn}_{t-8} + \hat{\gamma}_{31} \mathit{IMPn}_{t-1} \\
&\quad + \cdots+ \hat{\gamma}_{38} \mathit{IMPn}_{t-8} + \hat{\delta}_{31}
\mathit{ACTEn}_{t-1} + \cdots+ \hat{\delta}_{38} \mathit{ACTEn}_{t-8},
\end{aligned}
\end{cases}
\end{equation}
\\
where the letter $n$ in the variable name indicates that we are working
with national data.

\begin{figure}[t]
\includegraphics{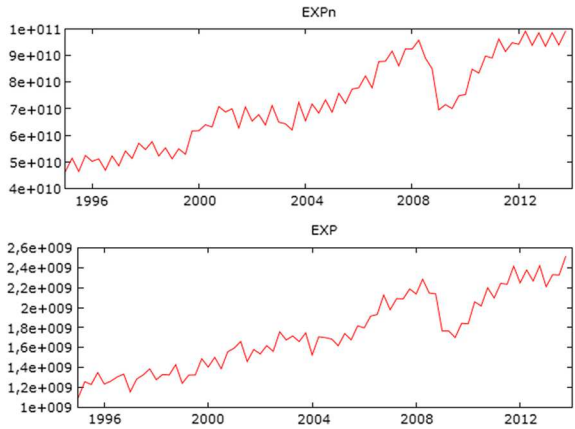}
\caption{Comparison between ${\mathit{EXPn}_t}$ and
${\mathit{EXP}_t}$}\label{7}
\end{figure}

\begin{figure}[t]
\includegraphics{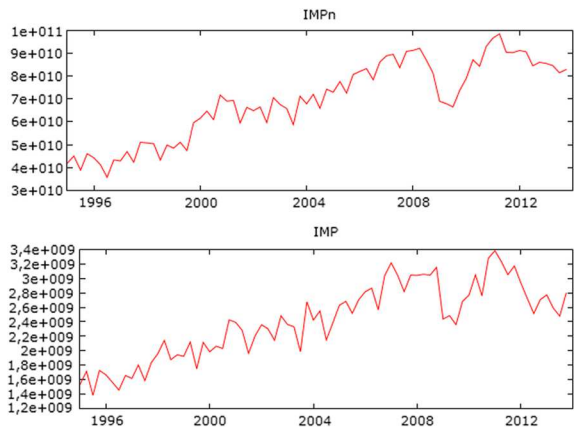}
\caption{Comparison of ${\mathit{IMPn}_t}$ and
${\mathit{IMP}_t}$}\label{8}
\end{figure}

\begin{figure}[t]
\includegraphics{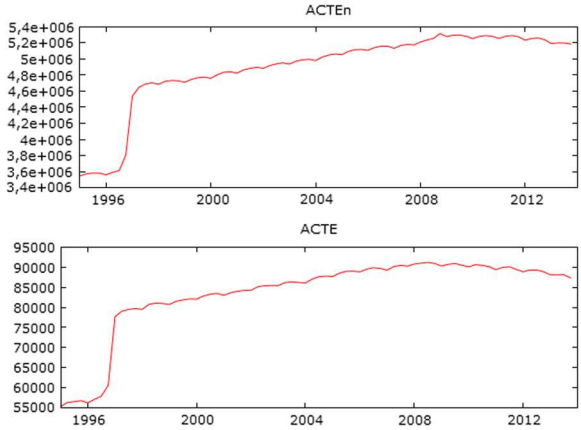}
\caption{Comparison of ${\mathit{ACTEn}_t}$ and
${\mathit{ACTE}_t}$}\label{9}\vspace*{-9pt}
\end{figure}

The $\mathit{Adjusted}\ R^2$ of the three equations in system \eqref{VAR(8)} are
respectively 0.95, 0.96, and 0.98. So this is a good VAR model; in
fact, Granger causality tests for \eqref{VAR(8)} present all $p$-values
of the F-statistic less than 0.01. So all the three variables can be
used to explain the others. In Figs.~\ref{7}, \ref{8}, and \ref{9}, we
note the extreme similarity of the provincial and national time series.
If we perform an~EG-ADF test to verify if this three couples of
variables are cointegrated, then we obtain that only the variables
$\mathit{ACTEn}_t$ and $\mathit{ACTE}_t$ are cointegrated with cointegrating coefficient
$\theta= 49.4948$. By comparing the correlation between a variable of
national data and the corresponding variables with provincial data we
note a high correlation level, even taking into account the provincial
variable delays. Below we present the correlation between $\mathit{EXPn}$ and
the delays of $\mathit{EXP}$:

\begin{center}
{\footnotesize
\begin{tabular}{cc}
\hline
${p}$ & ${\mathit{corr}(\mathit{EXPn}_t ; \mathit{EXP}_{t+p})}$ \\
\hline
--4 & 0.7918 \\

--3 & 0.8083 \\

--2 & 0.8985 \\

--1 & 0.9036 \\

0 & 0.9823 \\

1 & 0.8880 \\

2 & 0.8677 \\

3 & 0.7711 \\

4 & 0.7557 \\
\hline
\end{tabular}}\vspace*{3pt}
\end{center}
Then we have the correlation between $\mathit{IMPn}$ and the delays of $\mathit{IMP}$
\begin{center}
{\footnotesize
\begin{tabular}{cc}
\hline
${p}$ & ${\mathit{corr}(\mathit{IMPn}_t ; \mathit{IMP}_{t+p})}$ \\
\hline
--4 & 0.7490 \\

--3 & 0.7645 \\

--2 & 0.8428 \\

--1 & 0.8745 \\

0 & 0.9641 \\

1 & 0.8780 \\

2 & 0.8518 \\

3 & 0.7887 \\

4 & 0.7948 \\
\hline
\end{tabular}}\vspace*{3pt}
\end{center}
whereas the correlation between $\mathit{ACTEn}$ and the delays of $\mathit{ACTE}$ are
given by
\begin{center}
{\footnotesize
\begin{tabular}{cc}
\hline
${p}$ & ${\mathit{corr}(\mathit{IMPn}_t ; \mathit{IMP}_{t+p})}$ \\
\hline
--4 & \xch{0.6493}{0,6493} \\

--3 & \xch{0.7400}{0,7400} \\

--2 & \xch{0.8290}{0,8290} \\

--1 & \xch{0.9162}{0,9162} \\

0 & \xch{0.9947}{0,9947}\\

1 & \xch{0.9257}{0,9257} \\

2 & \xch{0.8464}{0,8464} \\

3 & \xch{0.7634}{0,7634} \\

4 & \xch{0.6771}{0,6771} \\
\hline
\end{tabular}}
\end{center}

\begin{figure}[t]
\includegraphics{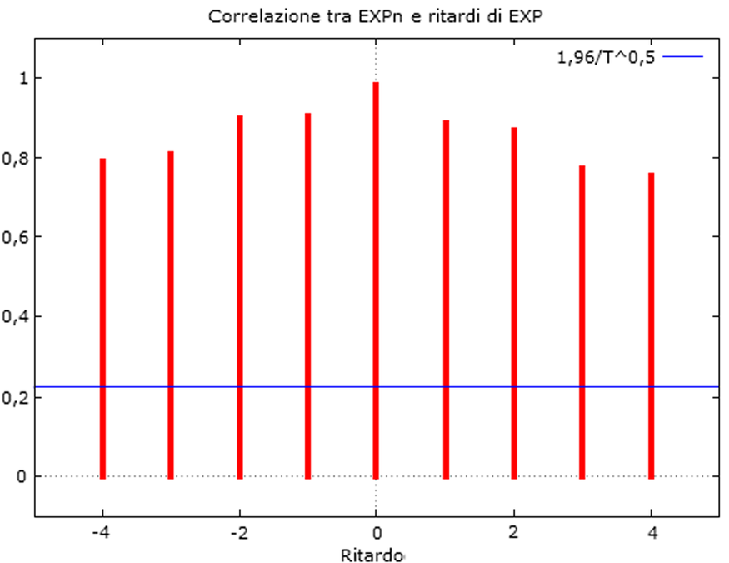}
\caption{Correlation between ${\mathit{EXPn}}$ and
${\mathit{EXP}}$}\label{10}
\end{figure}

\begin{figure}[t]
\includegraphics{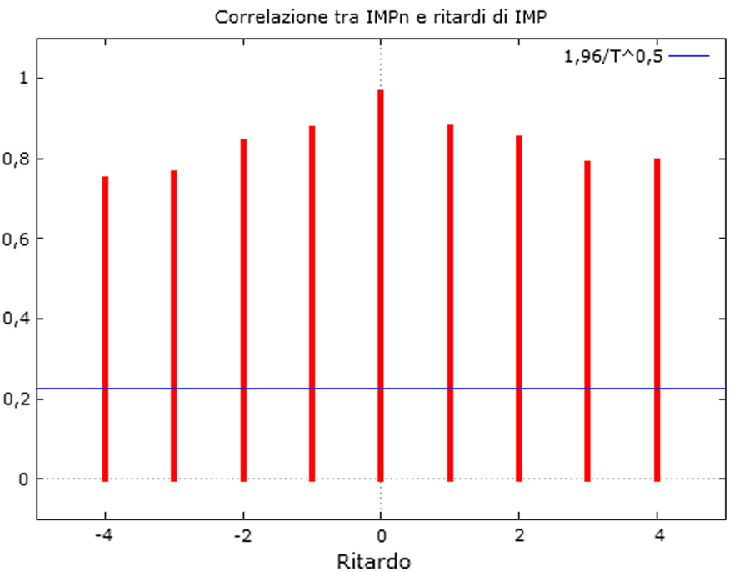}
\caption{Correlation between ${\mathit{IMPn}}$ and
${\mathit{IMP}}$}\label{11}
\end{figure}

\begin{figure}[t]
\includegraphics{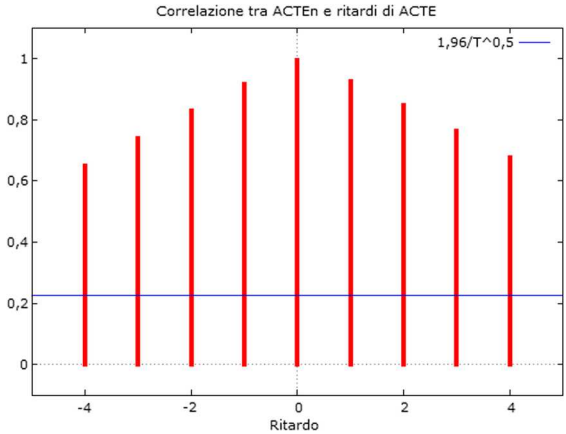}
\caption{Correlation between ${\mathit{ACTEn}}$ and
${\mathit{ACTE}}$}\label{12}
\end{figure}

Figures \ref{10}, \ref{11}, and \ref{12} show the correlation diagram
related to the national and provincial variables. We notice very high
values, which show the strong connection between what happens at the
national and the provincial levels.

\section{Conclusion}
We have presented an analysis of relevant time series related to the
import and export data concerning the Province of Verona, together with
a forecast analysis of the 2014 trend. Exploited techniques have been
treated in our first paper, and these two articles together constitute
a unitary project.
In this second part, we have paid attention to the quantitative
influence that certain macro economical events may have on considered
time series. In particular, we extrapolated three particularly
significant moments, namely the 2007--2008 world financial economic
crisis, with consequent decrease of import--export, a~break in 1995
probably due to the devaluation of the Lira, which did not cause a
decrease of the import, but resulted in an increase in exports of
Verona, and the vertical growth of the {\it Active enterprises}
parameter during 1995--1998, which has been caused by a change in the
related provincial regulation. It is worth to underline how our
analysis shows, by obtained numerical forecasts, a concrete possibility
for a partial recovery from the present economic crisis, especially
when taking into account the first quarters of 2014 and particularly
with regard to exports. The results obtained can be used for concrete
actions aimed, for example, to the optimization of territory economic
resources, even if a concrete economical program needs of a deeper
treatment for which, however, our analysis constitutes a rigorous and
effective basis. Concerning the latter, possible extensions may be
focused on analyzing import and export time series of specific products
to underline in which areas Verona is more specialized; then such
results could be used to understand where to invest more. Moreover, we
could perform a comparison analysis with analogous data belonging to
other cities of similar economical size, both in Italy and within the
European Community.

\section*{Acknowledgements}
The authors would like to thanks the {\it Camera di Commercio di
Verona} for the precious database that has been put at our disposal.
Any effective statistical/econometric analysis cannot be realized
without using real data. Any concrete forecast cannot be possible
without counting on such a kind of really precious time series.
Therefore, the present project would have not see the light without the
concrete help of the {\it Camera}. A particular acknowledgment goes to
Dr. Stefania Crozzoletti and to Dr. Riccardo Borghero.


\end{document}